\documentclass[%
superscriptaddress,
showpacs,
amsmath,amssymb,
prc
floatfix, ]%
{revtex4-1}
\usepackage{color}
\definecolor{mygray}{gray}{.9}
\definecolor{mypink}{rgb}{.99,.91,.95}
\definecolor{mycyan}{cmyk}{.3,0,0,0}
\usepackage{graphicx}% Include figure files
\usepackage{dcolumn}% Align table columns on decimal point
\usepackage{bm}% bold math
\newcommand{\element}[2]{~${}^{#1}${#2}}
\begin{document}

%\begin{CJK*}{GBK}{}

\title{Insight into nuclear midshell structures by study of $K$-isomers in rare-earth neutron-rich nuclei}

\author{Xiao-Tao He}% 
\email{hext@nuaa.edu.cn}
 \affiliation{College of Material Science and Technology, Nanjing University of Aeronautics and Astronautics, Nanjing 210016, China}
\author{Yu-Chun Li}%
 \affiliation{College of Material Science and Technology, Nanjing University of Aeronautics and Astronautics, Nanjing 210016, China}

\date{\today}

\begin{abstract}
Inspired by the newly discovered isomeric states in the rare-earth neutron-rich nuclei, high-$K$ isomeric states in neutron-rich samarium and gadolinium isotopes are investigated within the framework of the cranked shell model (CSM) with pairing correlation treated by a particle-number-conserving (PNC) method. The experimental multi-particle state energies and moments of inertia are reproduced quite well by the PNC-CSM calculations. A remarkable effect from the high-order deformation $\varepsilon_{6}$ is demonstrated. Based on the occupation probabilities, the configurations are assigned to the observed high-$K$ isomeric states. The lower $5^-$ isomeric state in $^{158}$Sm is preferred as the two-proton state with configuration $\pi\frac{5}{2}^{+}[413]\otimes\pi\frac{5}{2}^{-}[532]$. More low-lying two-particle states are predicted. The systematics of the electronic quadrupole transition probabilities, $B(E2)$ values along the neodymium, samarium, gadolinium and dysprosium isotopes and $N=96,98,100,102$ isotones chains is investigated to reveal the midshell collectivities. 
\end{abstract}

%\pacs{21.60.-n; 21.60.Cs; 23.20.Lv; 27.90.+b}%

%21.10.-k 	Properties of nuclei; nuclear energy levels
%21.60.-n 	Nuclear structure models and methods
%21.60.Cs Shell model
%23.20.Lv transitions and level energies
%27.90.+b A 220
\maketitle

\section{Introduction}

The rare-earth neutron-rich nuclei lie in the midshell region between the closed shells of proton $Z=50$, $82$ and neutron $N=82$, $126$. The high spin spectroscopy of these nuclei can provide important insights into the midshell collectivity, changes in nuclear shape and deformed sub-shells in a less-explored single-particle spectrum region. Furthermore, the pygmy rare-earth peak at $A\sim160$ of the $r-$process abundance is believed to arise from strong midshell nuclear deformation~\cite{SurmanR1997_PRL79,MumpowerM2012_PRC85}. The nuclear structure inputs of the rare-earth nuclei can lead to an improved understanding of the $r-$process nucleosynthesis~\cite{MumpowerM2017_JoPGNaPP44}. However, due to their neutron excess, detailed structure informations are very difficult to be revealed by the experiment in these rare-earth neutron-rich nuclei. 

The recent experimental progresses in the neutron-rich $A=150-170$ region~\cite{PatelZ2017_PRC96,YokoyamaR2017_PRC95,IdeguchiE2016_PRC94,SoederstroemP2016_PLB762,PatelZ2016_PLB753,PatelZ2016_EWoC123,WatanabeH2016_PLB760,PatelZ2014_PRL113,SimpsonG2009_PRC80,UrbanW2009_PRC80} are attributed to a great extent to the existence of the nuclear isomeric state.  A nucleus can be "traped" in an aligned spin orientation relative to its symmetric axis to form the $K$ isomeric state (or $K$-isomer), where $K$ is a quantum number representing the projection of the total nuclear spin along the symmetry axis of the nucleus. $K$-isomer arises from the multi-particle state, of which the transition to a lower energy state with a different $K$ value is inhibited by the $\Delta K\leq\lambda$ selection rule where $\lambda$ is the multipole order of the transition. However, symmetry-breaking processes make these transitions possible to process with the $\Delta K-\lambda$ related hindrance factor~\cite{WalkerP1999_N399,WalkerP2015_PS91}. The multi-particle state is formed by breaking pairs of nucleon. The excitation energy and configuration of multi-particle state depend strongly on the position of the specific single-particle orbitals near the Fermi surface and correlations, such as pairing. $K$-isomer appears only in axially symmetric deformed nuclei well away from the closed shell. This implies that $K$-isomer cannot be a pure intrinsic state without considering the collective rotation. As it has been observed in the experiment, $K$-isomer, especially based on lower-energy intrinsic state of less complexity, is often associated with a rotational band, of which it forms the bandhead. However, assigning the $K$ value as the bandhead of a rotational band is not strict due to the $K$- mixing induced by Coriolis. The rotational band properties are affected strongly by the intrinsic states, especially when the components of the configuration include the high-$j$ low-$\Omega$ orbitals, and in turn it can be used to test the proposed configuration. Investigations of $K$-isomers have been reviewed recently in Refs.~\cite{DracoulisG2016_RoPiP79,KondevF2015_ADaNDT103_104,WalkerP2015_PS91,WalkerP2001_HI135,WalkerP1999_N399}.   

An impressive progress has been achieved in experimental studies of samarium and gadolinium isotopes very recently~\cite{PatelZ2017_PRC96,YokoyamaR2017_PRC95,PatelZ2016_PLB753,PatelZ2016_EWoC123,PatelZ2014_PRL113,SimpsonG2009_PRC80}. The most exiting discoveries include: $^{160}$Sm became the lightest nucleus with a known four-quasiparticle $K$-isomer~\cite{PatelZ2016_PLB753}, $K$-isomers were firstly observed in very neutron-rich nuclei $^{164}$Sm and $^{166}$Gd~\cite{PatelZ2014_PRL113} and in odd-A samarium isotopes~\cite{PatelZ2017_PRC96,YokoyamaR2017_PRC95}. Besides the important structure informations mentioned above, these experimental data will provide a good test ground of the present nuclear theories in the most unknown territory in the nuclear landscape. However, there are still no detailed theoretical studies being performed according to these observations. 

In the present work, the newly observed $K$-isomers and their associated rotational bands in samarium and gadolinium neutron-rich isotopes are investigated by the cranked shell model with pairing treated by the particle-number conserving method. This is the first time for the PNC-CSM calculations being performed on such neutron-rich nuclear mass region. The experimental multi-particle state energies and moments of inertia can be reproduced quite well. Configurations are assigned to the observed $K$-isomers. More low-lying exited states are predicted and the collectivity among the $Z=60-66$ isotopes and $N=96-102$ isotones are discussed in details.  

\section{Theoretical Framework}{\label{Sec:PNC}}
In the frame of cranked shell model, a particle-number conserving method is used to treat the pairing correlations. In this method, the cranked shell model Hamiltonian is diagonalized directly in a truncated Fock space and a pair-broken excited configuration is defined by blocking the real particles in the single-particle orbitals~\cite{ZengJ1994_PRC50a,ZengJ1983_NPA405}. By this way, particle number is conserved and the Pauli blocking effect can be treated spontaneously~\cite{ZengJ1983_NPA405,WuC1989_PRC39,ZengJ1994_PRC50,XinX2000_PRC62}. The PNC-CSM method has previously been applied successfully to describe the properties of the normal deformed nuclei in $A=170$ mass region~\cite{ZengJ1994_PRC50,WuC1991_PRC44,ZengJ2002_PRC65,LiuS2002_PRC66,LiuS2004_NPA735,ZengJ2001_PRC63}, superdeformed nuclei in $A=150, 190$ mass region~\cite{WuC1992_PRC45,LiuS2002_PRC66a,LiuS2004_NPA736,ZengJ1991_PRC44,HeX2005_NPA760} and most recently in the light $Z\approx N$ nuclei around $Z=40$ mass region~\cite{XiangX2018_CPC42}, the structures of the heaviest actinides and light superheavy nuclei around $Z=100$ mass region~\cite{LiY2016_SCPMA59, ZhangZ2016_SCPMA59, ZhangZ2013_PRC87, ZhangZ2012_PRC85, ZhangZ2011_PRC83, HeX2009_NPA817}, and the high-$K$ isomeric states in the rare-earth and actinide nuclei~\cite{LiB2013_CPC37, ZhangZ2009_PRC80, ZhangZ2009_NPA816}. Recent years, the PNC method has been implemented to other nuclear theoretical models to treat the pairing correlation successfully. One example is to the total-routhian-surface (TRS) model, which has been used to study the high-$K$ multi-particle states in $^{178}$W~\cite{FuX2013_PRC87} and Hf isotopes~\cite{LiangW2015_PRC92} and the $8^-$ isomers in $N=150$ light superheavy nuclei~\cite{FuX2014_PRC89}. Another example is to the covariant density functional theory, in which PNC method, referred to as Shell-model-like approach (SLAP), is applied to treat the cranking many-body Hamiltonian to investigate the rotational structures in $^{60}$Fe~\cite{ShiZ2018_PRC97}.

The CSM Hamiltonian in the rotating frame reads,
\begin{eqnarray}
\label{eq:H_CSM}
 H_\mathrm{CSM}
  & = & H_{\rm SP}-\omega J_x + H_\mathrm{P}(0)+H_\mathrm{P}(2)\\
 \nonumber & = & \sum_{\xi}h_{\xi}-\omega J_{x}-G_{0}\sum_{\xi \eta }
 a_{\xi }^{\dagger}a_{\overline{\xi }}^{\dagger }a_{\overline{\eta }}a_{\eta }
 -G_{2}\sum_{\xi \eta } q_{2}(\xi) q_{2}(\eta)
 a_{\xi}^{\dagger } a_{\overline{\xi}}^{\dagger}
 a_{\overline{\eta}}a_{\eta}\ ,
\end{eqnarray}
where $h_\xi$ is the single-particle Hamiltonian with an one-body potential (here the Nilsson potential is adopted), $-\omega J_x$ is the Coriolis interaction with the cranking frequency $\omega$ about the $x$ axis (perpendicular to the nuclear symmetry z axis). $H_\mathrm{P}(0)$ and $H_\mathrm{P}(2)$ are the monopole- and quadrupole-pairing correlations, respectively. $\xi$ ($\eta$) is the eigen state of the single-particle Hamiltonian $h_{\xi}$, and $\bar{\xi}$ ($\bar{\eta}$) denotes its time-reversed state. $a_{\xi }^{\dagger}a_{\overline{\xi }}^{\dagger }\equiv s_\xi^\dagger$ $(a_{\bar{\eta}} a_\eta\equiv s_\eta) $ is the time reversal pair creation (annihilation) operator with $a_{-\xi}^{\dagger}=(-1)^{\Omega-1/2}a_{\overline{\xi}}^{\dagger }$, where $\Omega$ is the z component of the single-particle angular momentum. $q_{2}(\xi) = \sqrt{{16\pi}/{5}}\langle \xi |r^{2}Y_{20} | \xi \rangle$ is the diagonal element of the stretched quadrupole operator. 

In the rotating frame, the symmetry of the time reversal is broken while the symmetry of rotation by $\pi$ around the $x$ axis, $R_{x}(\pi)=e^{-i\pi\alpha}$, is retained. Signature $\alpha=\pm1/2$ are the eigenvalues of $R_{x}(\pi)$. The time reversal representation can be transformed to the signature basis by constructing the simultaneous eigenstates of $\{h_0(\omega)=h_\xi-\omega j_{x},j_{z}^2,R_{x}(\pi)\}$ as $| \xi\alpha\rangle=\frac{1}{\sqrt{2}}[1-e^{-i\pi\alpha}R_{x}(\pi)]|\xi\rangle$. 
 
In the signature representation, pairing is expressed as,
\begin{equation}
H_P=-G\sum_{\xi\eta>0}(-1)^{\Omega_\xi-\Omega_\eta}\beta_{\xi+}^{\dagger}\beta_{\xi-}^{\dagger}\beta_{\eta-}\beta_{\eta+},
\end{equation}
where $\beta_{\xi\alpha=\pm1/2}^{\dagger}=\frac{1}{\sqrt2}\left[a_{\xi}^{\dagger}\pm(-1)^{N_{\xi}}a_{-\xi}^{\dagger}\right]$ is the \textit{real} particle creation operator of the state $|\xi\alpha\rangle$. The eigenstates of the cranked single-particle Hamiltonian is obtained by diagonalizing $h_0(\omega)=h_\xi-\omega j_{x}$ in the signature $|\xi\alpha\rangle$ space as,
\begin{equation}
|\mu\alpha\rangle=\sum_\xi C_{\mu\xi}(\alpha)|\xi\alpha\rangle, \ \ \ \ [C_{\mu\xi}(\alpha) \rm {\ is\ real}],
\end{equation}
which is characterized by parity $\pi$ and signature $\alpha$. The cranked Nilsson levels are given by the single-particle energy eigenvalues $\epsilon_{\mu\alpha}$ versus frequency $\hbar\omega$. Hereafter, $|\mu\alpha\rangle $ is briefly denoted by $|\mu\rangle$. A cranked many-particle configuration (CMPC) of a $n$-particle system is expressed as,
\begin{equation}\label{eq:CMPC}
|i\rangle=|\mu_1\mu_2\cdots\mu_n \rangle=b_{\mu_1}^{\dagger}b_{\mu_2}^{\dagger}\cdots b_{\mu_n}^{\dagger}|0\rangle
\end{equation}
where $b_{\mu\pm}^{\dagger}=\sum_\xi C_{\mu\xi}(\pm)\beta_{\xi\pm}^{\dagger}$ is the \textit{real} particle creation operator of the cranked state $|\mu\rangle$. Each configuration $|i\rangle$ is characterized by the particle-number $n$, parity $\pi$, signature $\alpha$ and seniority $\nu$ (number of unpaired particles). 

In the cranked basis, the one-body part of $H_{\rm CSM}$ is,
\begin{equation}
H_0=\sum_{\mu\alpha}\epsilon_{\mu\alpha}b_{\mu\alpha}^{\dagger}b_{\mu\alpha},
\end{equation}
and the pairing reads,
\begin{eqnarray}
H_P(0)&=&-G_0\sum_{\mu\mu{\prime}\nu\nu{\prime}}f^{\ast}_{\mu\mu{\prime}}f_{\nu{\prime}\nu}
b_{\mu+}^{\dagger}b_{\mu{\prime}-}^{\dagger}b_{\nu-}b_{\nu{\prime}+},\\
\nonumber f^{\ast}_{\mu\mu{\prime}}&=&\sum_{\xi>0}(-)^{\Omega_\xi}C_{\mu\xi}(+)C_{\mu{\prime}\xi}(-), \\ 
\nonumber f_{\nu{\prime}\nu}&=&\sum_{\eta>0}(-)^{\Omega_\eta}C_{\nu{\prime}\eta}(+)C_{\nu\eta}(-).
\end{eqnarray}
and,
\begin{eqnarray}
H_P(2)&=&-G_2\sum_{\mu\mu{\prime}\nu\nu{\prime}}g^{\ast}_{\mu\mu{\prime}}g_{\nu{\prime}\nu}
b_{\mu+}^{\dagger}b_{\mu{\prime}-}^{\dagger}b_{\nu-}b_{\nu{\prime}+},\\
\nonumber g^{\ast}_{\mu\mu{\prime}}&=&\sum_{\xi>0}(-)^{\Omega_\xi}C_{\mu\xi}(+)C_{\mu{\prime}\xi}(-)q_2(\xi), \\
\nonumber g_{\nu{\prime}\nu}&=&\sum_{\eta>0}(-)^{\Omega_\eta}C_{\nu{\prime}\eta}(+)C_{\nu\eta}(-)q_2(\eta).
\end{eqnarray}

The $H_{\rm CSM}$ is diagonalized in a sufficiently large cranked many-particle configuration space, which is constructed by including the configurations with energy $E_i\le E_0+ E_c$, where $E_0$ is the energy of the lowest configuration and $E_c$ is the cutoff energy.  The eigenstate of $H_{\rm CSM}$ reads,

\begin{equation}\label{eq:eigenstate}
 | \psi \rangle = \sum_{i} C_i | i \rangle, ~~ C_i \text{ is real,}
\end{equation}
This converged solution can always be obtained even for a pair-broken state. The Pauli blocking effect is treated spontaneously while it does not in the BCS or Hartree-Fock-Bogoliubov quasi-particle (qp) formalism. For the seniority $\nu=0$ ground state ($K^\pi=0^+$) of an even-even nucleus (qp vacuum in the BCS formalism), each $| i \rangle$ in Eq.\ref{eq:eigenstate} is of the form~\cite{WuX2011_PRC83},
\begin{equation}
|i\rangle=|\mu_1\bar{\mu}_1\cdots\mu_{k}\bar{\mu}_{k} \rangle=b_{\mu_1}^{\dagger}b_{\bar{\mu}_1}^{\dagger}\cdots b_{\mu_k}^{\dagger}b_{\bar{\mu}_k}^{\dagger}|0\rangle
\end{equation}
where $k=(n-\nu)/2$. For the seniority $\nu=1$ state in an odd-even nucleus, $| i \rangle$ is of the form,
\begin{eqnarray} 
|i\rangle=|\sigma_{1}\mu_1\bar{\mu}_1\cdots\mu_{k}\bar{\mu}_{k} \rangle=b_{\sigma_1}^{\dagger}b_{\mu_1}^{\dagger}b_{\bar{\mu}_1}^{\dagger}\cdots b_{\mu_k}^{\dagger}b_{\bar{\mu}_k}^{\dagger}|0\rangle,\ \ \ \ (\sigma \neq \mu),
\end{eqnarray}
where $\sigma_1$ is the blocked single-particle state. The angular momentum projection along the nuclear symmetry $z$-axis $K=\Omega_{\sigma_1}$, $\pi=(-)^{N_{\sigma_1}}$. For a pairing-broken state in an even-even nucleus with the seniority $\nu=2$, $| i \rangle$ is of the form,
\begin{eqnarray}
|i\rangle=|\sigma_1\sigma_2\mu_1\bar{\mu}_1\cdots\mu_{k}\bar{\mu}_{k} \rangle=b_{\sigma_1}^{\dagger}b_{\sigma_2}^{\dagger}b_{\mu_1}^{\dagger}b_{\bar{\mu}_1}^{\dagger}\cdots b_{\mu_k}^{\dagger}b_{\bar{\mu}_k}^{\dagger}|0\rangle,\ \ \ \ (\sigma \neq \mu),
\end{eqnarray}
where $\sigma_1$, $\sigma_2$ are the two blocked single-particle states. Note that for two given blocked levels, different occupation of $(\sigma_1\bar{\sigma}_2)$, $(\bar{\sigma}_1\sigma_2)$ and $(\bar{\sigma}_1\bar{\sigma}_2)$ are considered as well. This leads to four sequences, i.e. $K=|\Omega_{\sigma_1}\pm\Omega_{\sigma_2}|$ combined with $\alpha=0,1$. The parity of these configurations is $\pi=(-)^{N_{\sigma_1}+N_{\sigma_2}}$. The configurations of higher-seniority $\nu>2$ states are similarly constructed, and the diagonalisation remains the same. 

As a matter of fact, when $\omega\neq0$, $\nu$ and $K$ are not exactly good quantum numbers due to the Coriolis interaction. Some forms of $K$-mixing exist to enable the $K$-forbidden transition observed in an axially symmetric nucleus with many low-lying rotational bands. Nevertheless, at the low-$\omega$ region, $\nu$ and $K$ may still be served as useful quantum numbers characterizing a low-lying excited rotational band.

\section{Results and discussions}{\label{Sec:results}}
\subsection{Nilsson single-particle levels}

The single-particle levels are particular important to the low-lying multi-particle states. In the present calculations, Nilsson states are calculated within the valence single-particle space of proton $N =0\sim5$ and neutron $N=0\sim6$ major shells. The Nilsson parameters ($\kappa, \mu$) are taken from the Lund systematics~\cite{NilssonS1969_NPA131}. The deformation parameters $\varepsilon_2$, $\varepsilon_4$ and $\varepsilon_6$ are taken from the table of M\"oller \textit{et al.} in 1995~\cite{MoellerP1995_ADaNDT59} for all the nuclei studied here except for the value of $\varepsilon_{6}$ for $^{164}$Sm, which is adopted from the new table of M\"oller \textit{et al.} in 2016~\cite{MoellerP2016_ADNDT109-110}.   

\begin{figure}%[!t]
   \includegraphics[scale=0.35]{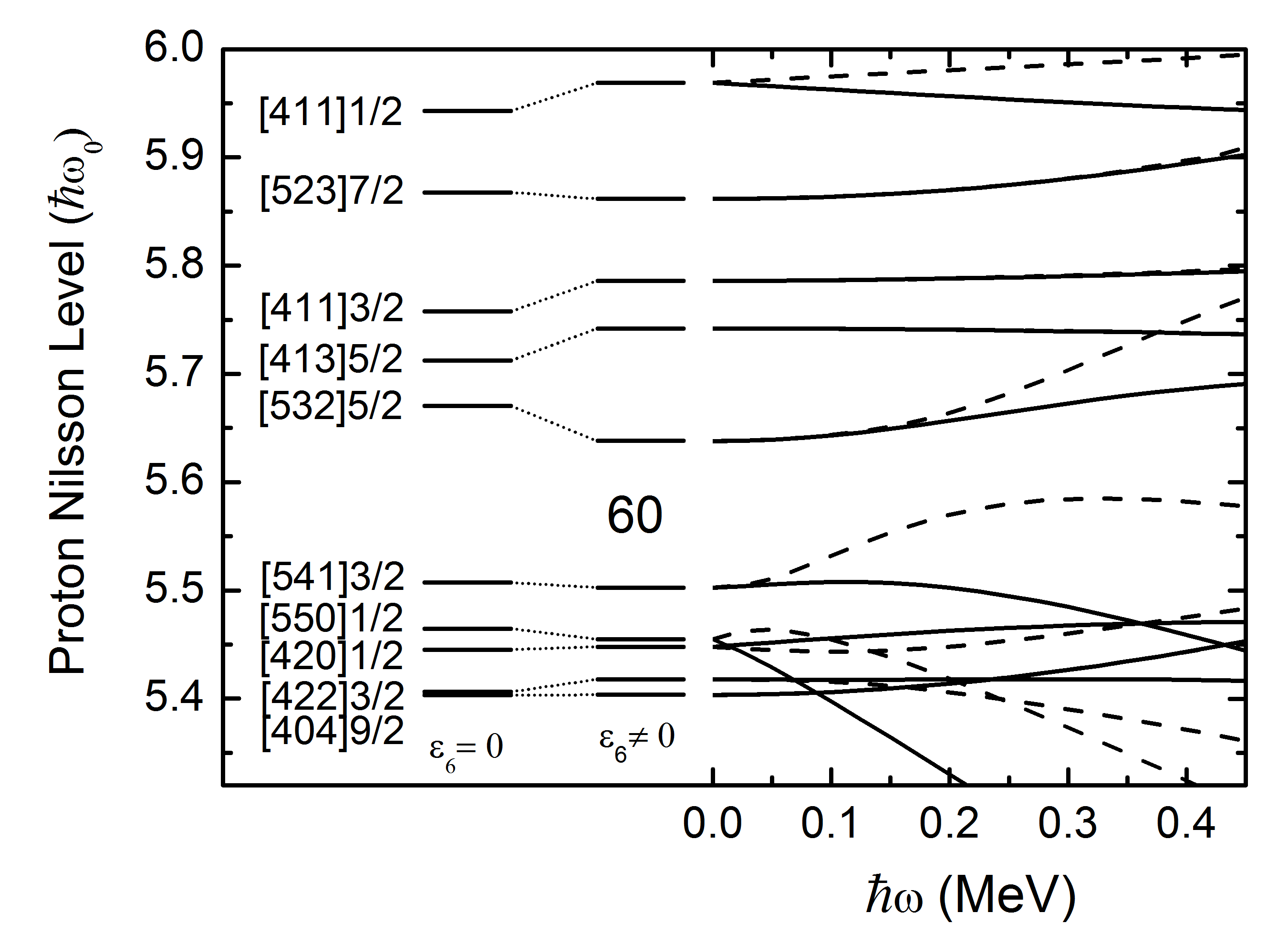}
    \includegraphics[scale=0.35]{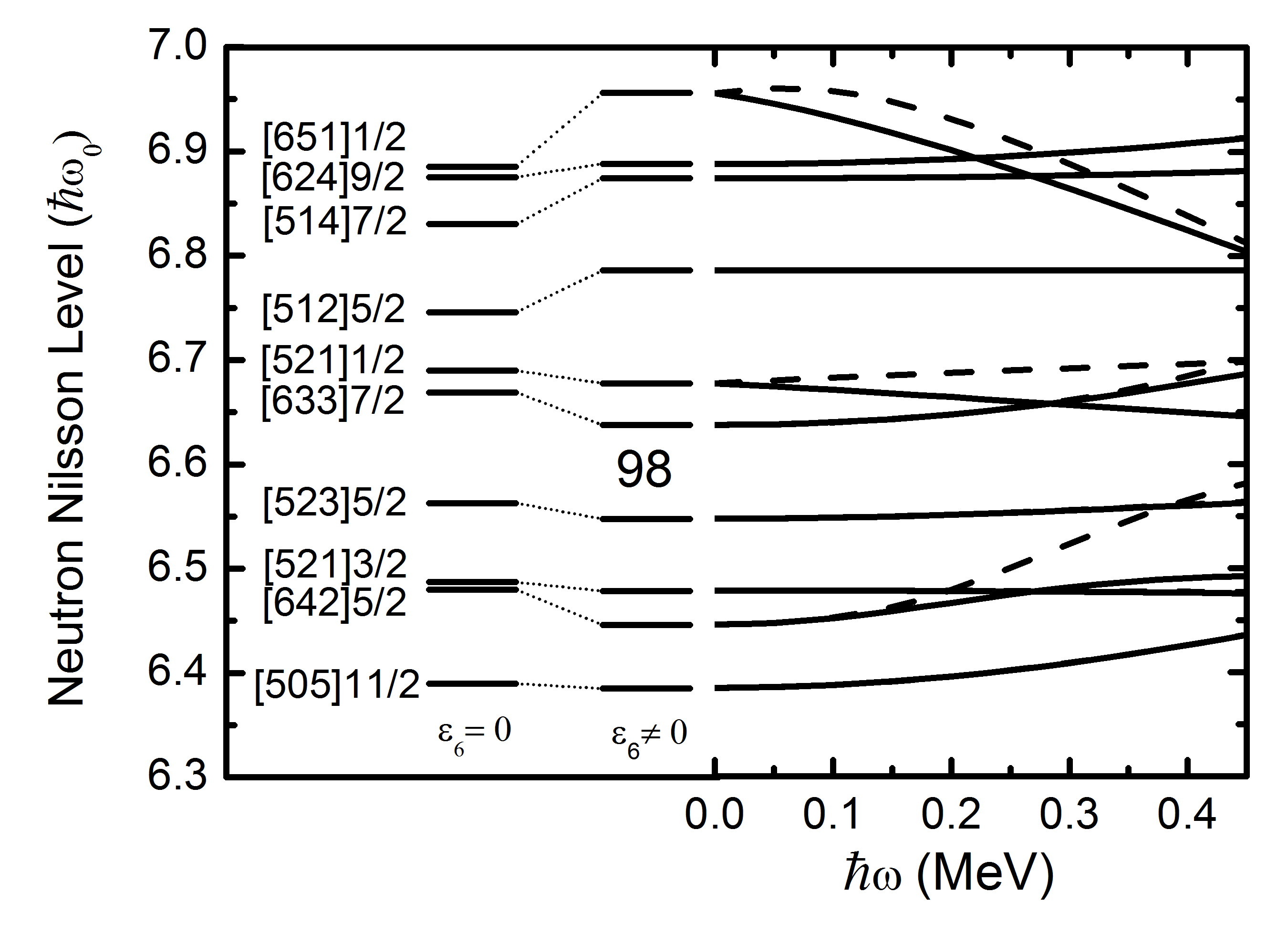}
    \caption{\label{fig:Nilsson}
Cranked Nilsson levels near the Fermi surface of \element{160}{Sm} for proton (left) and neutron (right) with signature $\alpha=+1/2$ (solid) and $\alpha=-1/2$ (dashed). }
\end{figure}

The high-order axial deformation $\varepsilon_{6}$ is included due to its important effect on the deformed shell gaps at $Z=60,62$ and $N=98,102$, and the further influences on the multi-particle excitation energy and the moment of inertia in the rare-earth neutron-rich region. The Nilsson levels with non-zero $\varepsilon_{6}$ at rotational frequency $\hbar\omega=0$ are compared with the results of $\varepsilon_{6}=0$ (keep $\varepsilon_{2}$ and $\varepsilon_{4}$ the same) for $^{160}$Sm in Fig.~\ref{fig:Nilsson}. The single-particle level structures of other nuclei studied in the present work are similar. Therefore they are not shown here. For proton, compared to the $\varepsilon_{6}=0$ case, including of nonzero $\varepsilon_{6}$ leads to a less pronounced energy gap at $Z=60$. A new energy gap at $Z=62$ arises while the one at $Z=66$ disappears. Based on these effects, the low-lying multi-proton state energies would be increased for samarium isotopes. For neutron, by including the nonzero $\varepsilon_{6}$, the energy gap at $N=98$ is depressed and the one at $N=102$ appears. 

Potential energy surface calculations demonstrate that the $\varepsilon_{6}$ effect on the multi-particle state is important. It leads to $50-250$ keV variations compared to $\varepsilon_{6}=0$ calculations in $^{160,164}$Sm and $^{166}$Gd~\cite{PatelZ2014_PRL113, PatelZ2016_PLB753}. Nevertheless, high-order deformation influences are still very intricate, especially for neutrons in the neutron-rich region, in which the knowledge of single-particle level structure is very limited. Moreover, the value of $\varepsilon_{6}$ is strongly model dependent. Therefore, more deep and comprehensive investigations of the $\varepsilon_{6}$ deformation effect on the single-particle levels in this region are urgent, which is beyond present work.          
   
%%%%%%%%%%%%%%%%%%%%%%%%%%%%%%%%%%%%%%%%%%%%%
%%%%%%%%%%%%%%%%%%%%%%%%%%%%%%%%%%%%%%%%%%%%%

\subsection{Pairing parameters}

The effective pairing strengths $G_0$ and $G_2$ can be determined by the odd-even differences in nuclear binding energies in principle. For the rare-earth neutron-rich nuclei, due to lack of experimental data, their values are fitted by moment of inertia. The effective pairing strengths are connected with the dimensions of the truncated CMPC space. In the present calculations, the CMPC spaces for all the nuclei involved are constructed in the proton $N=3,4,5$ and neutron $N=4,5,6$ shells. The dimensions of the CMPC space are about 700 for both of proton and neutron. The corresponding effective pairing strengths are $G_{0p}=0.20{\rm MeV}$, $G_{2p}=0.02{\rm MeV}$ and $G_{0n}=0.23{\rm MeV}$, $G_{2n}=0.02{\rm MeV}$ for proton and neutron, respectively. The stability of the PNC-CSM calculation against the variation of the dimensions of the CMPC space has been investigated in Refs.~\cite{ZengJ1983_NPA405,LiuS2002_PRC66,ZhangZ2012_PRC85}. For the yrast and low-lying excited states, the number of important CMPC (weight $>10^{-2}$) is very limited ($<20$). In the present calculations, almost all of CMPC with weight $>10^{-3}$ are taken into account. 

\subsection{Band head}

%%%%%%%%%%%%%%%%%%%%%%%%%%%%%%%%%%%%%%%%%%%%%
\begin{table}
\caption{\label{tab:Sm_K_isomer} Low-lying multi-particle states of samarium isotopes predicted by the PNC-CSM calculations. The excitation energies predicted by potential energy surface calculations (marked by letter a), by blocked-BCS calculations (marked by letter b) and by projected shell-model calculations (marked by letter c) are listed for comparison. The experimental data are taken from Refs.~\cite{PatelZ2017_PRC96,PatelZ2016_PLB753,PatelZ2014_PRL113,WangE2014_PRC90,SimpsonG2009_PRC80,nndc}. }
\begin{ruledtabular}
\begin{tabular}{ccclcr}
    $K^\pi$&Configuration&$E_x$(MeV)& $E_x^{a,b,c}$(MeV) & $E_{x}^{exp}$(MeV)\\ \hline
    $^{158}$Sm&&&&\\
    $5^-$   &$\pi\frac{5}{2}^{+}[413]\otimes\pi\frac{5}{2}^{-}[532]$   &1.227  & $\ \ \ \ \ \ \ \ \ \ \ \ \ \ \ \ \ \ \ 1.762^c$ &1.279 \\
    $4^-$   &$\pi\frac{3}{2}^{+}[411]\otimes\pi\frac{5}{2}^{-}[532]$   &1.506  &                  & \\
    $4^+$  &$\pi\frac{5}{2}^{+}[413]\otimes\pi\frac{3}{2}^{+}[411]$   &1.825  &                 & \\
    $6^+$ &$\pi\frac{7}{2}^{-}[523]\otimes\pi\frac{5}{2}^{-}[532]$    & 2.122  &                                       & \\
    $6^-$   &$\pi\frac{5}{2}^{+}[413]\otimes\pi\frac{7}{2}^{-}[523]$   & 2.419 &                 &\\
    $4^+$  &$\nu\frac{3}{2}^{-}[521]\otimes\nu\frac{5}{2}^{-}[523]$  &1.478  &                 &\\
    $5^-$   &$\nu\frac{5}{2}^{-}[523]\otimes\nu\frac{5}{2}^{+}[642]$ & 2.097 & $\ \ \ \ \ \ \ \ \ \ \ \ \ \ \ \ \ \ \ 1.441^c$ & 1.322\\
    $5^-$   &$\nu\frac{3}{2}^{-}[521]\otimes\nu\frac{7}{2}^{+}[633]$  & 2.165 & $             $ & \\
    $^{160}$Sm&&&&\\
    $5^-$   &$\pi\frac{5}{2}^{+}[413]\otimes\pi\frac{5}{2}^{-}[532]$   &1.276 & $1.457^a\ 1.032^b\ 1.791^c$ &1.361  \\
    $4^-$   &$\pi\frac{3}{2}^{+}[411]\otimes\pi\frac{5}{2}^{-}[532]$   &1.572 & $\ \ \ \ \ \ \ \ \ 1.569^b$                            &\\
    $4^+$  &$\pi\frac{5}{2}^{+}[413]\otimes\pi\frac{3}{2}^{+}[411]$   &1.890 & $\ \ \ \ \ \ \ \ \ 1.631^b$                           & \\
    $6^+$ &$\pi\frac{7}{2}^{-}[523]\otimes\pi\frac{5}{2}^{-}[532]$    & 2.115  &                                       & \\
    $6^-$   &$\pi\frac{5}{2}^{+}[413]\otimes\pi\frac{7}{2}^{-}[523]$   &2.418 & $             $                           & \\
    $4^-$   &$\pi\frac{5}{2}^{+}[413]\otimes\pi\frac{3}{2}^{-}[541]$   &2.507 & $\ \ \ \ \ \ \ \ \ 2.107^b$                           & \\
    $6^-$   &$\nu\frac{5}{2}^{-}[523]\otimes\nu\frac{7}{2}^{+}[633]$  &1.508 &$1.727^a\ 1.401^b$            &1.468  \\
    $3^+$  &$\nu\frac{1}{2}^{-}[521]\otimes\nu\frac{5}{2}^{-}[523]$   &1.811  & $\ \ \ \ \ \ \ \ \ 1.674^b$                         & \\
    $5^-$   &$\nu\frac{3}{2}^{-}[521]\otimes\nu\frac{7}{2}^{+}[633]$   & 2.053 &                                           & \\
    $2^+$  &$\nu\frac{1}{2}^{-}[521]\otimes\nu\frac{3}{2}^{-}[521]$   & 2.345 &                                          & \\
    $4^-$   &$\nu\frac{1}{2}^{-}[521]\otimes\nu\frac{7}{2}^{+}[633]$   &2.408  &$\ \ \ \ \ \ \ \ \ \ \ \ \ \ \ \ \ \ \ 1.290^c$                         & \\
    $5^+$  &$\nu\frac{5}{2}^{-}[523]\otimes\nu\frac{5}{2}^{-}[512]$   &2.663  &$\ \ \ \ \ \ \ \ \ 1.953^b$                         & \\
    $11^+$&\{$\pi\frac{5}{2}^{+}[413]\otimes\pi\frac{5}{2}^{-}[532]$,  &2.784  &$3.214^a\ 2.433^b$          &2.757 \\
                &\ \ $\nu\frac{5}{2}^{-}[523]\otimes\nu\frac{7}{2}^{+}[633]$\}&&\\
    $8^-$   &\{$\pi\frac{5}{2}^{+}[413]\otimes\pi\frac{5}{2}^{-}[532]$,  &3.087 &$\ \ \ \ \ \ \ \ \ 2.706^b$                        &\\
                &\ \ $\nu\frac{1}{2}^{-}[521]\otimes\nu\frac{5}{2}^{-}[523]$\}  &&\\
    $10^+$&\{$\pi\frac{3}{2}^{+}[411]\otimes\pi\frac{5}{2}^{-}[532]$,   &3.080  &$\ \ \ \ \ \ \ \ \ 2.970^b$                        &\\
                &\ \ $\nu\frac{5}{2}^{-}[523]\otimes\nu\frac{7}{2}^{+}[633]$\}  &&\\
    $^{162}$Sm&&&&\\
    $5^-$  &$\pi\frac{5}{2}^{+}[413]\otimes\pi\frac{5}{2}^{-}[532]$    &1.321  &$\ \ \ \ \ \ \ \ \ 1.000^b\ 1.911^c$                         & \\
    $4^-$  &$\pi\frac{3}{2}^{+}[411]\otimes\pi\frac{5}{2}^{-}[532]$    &1.624  &$\ \ \ \ \ \ \ \ \ 1.547^b$                                        & \\
    $4^+$ &$\pi\frac{5}{2}^{+}[413]\otimes\pi\frac{3}{2}^{+}[411]$    &1.861  &$\ \ \ \ \ \ \ \ \ 1.614^b$                                       & \\
    $6^+$ &$\pi\frac{7}{2}^{-}[523]\otimes\pi\frac{5}{2}^{-}[532]$    & 2.108  &                                       & \\
    $4^-$  &$\pi\frac{5}{2}^{+}[413]\otimes\pi\frac{3}{2}^{-}[541]$    & 2.399 &                                        &\\
    $6^-$  &$\pi\frac{7}{2}^{-}[523]\otimes\pi\frac{5}{2}^{-}[413]$    & 2.438  &                                       & \\ 
    $4^-$  &$\nu\frac{1}{2}^{-}[521]\otimes\nu\frac{7}{2}^{+}[633]$  &1.155  &$\ \ \ \ \ \ \ \ \ 1.043^b\ 1.096^c$      &1.011\\
    $3^+$ &$\nu\frac{1}{2}^{-}[521]\otimes\nu\frac{5}{2}^{-}[523]$   &1.799  &                                        &\\
    $6^-$  &$\nu\frac{5}{2}^{-}[512]\otimes\nu\frac{7}{2}^{+}[633]$   &2.010   &$\ \ \ \ \ \ \ \ \ 1.797^b$ &\\
    $^{164}$Sm& &&&\\
    $5^-$  &$\pi\frac{5}{2}^{+}[413]\otimes\pi\frac{5}{2}^{-}[532]$    &1.210  &$1.411^a$                        &\\
    $4^-$  &$\pi\frac{3}{2}^{+}[411]\otimes\pi\frac{5}{2}^{-}[532]$    &1.560  &$1.907^a$                        & \\
    $4^+$ &$\pi\frac{5}{2}^{+}[413]\otimes\pi\frac{3}{2}^{+}[411]$    &1.633  &                                       & \\
    $6^+$ &$\pi\frac{7}{2}^{-}[523]\otimes\pi\frac{5}{2}^{-}[532]$    &2.024  &                                       & \\ 
    $6^-$  &$\pi\frac{7}{2}^{-}[523]\otimes\pi\frac{5}{2}^{-}[413]$    &2.099  &                                       & \\ 
    $4^-$  &$\pi\frac{5}{2}^{+}[413]\otimes\pi\frac{3}{2}^{-}[541]$    &2.213  &$2.195^a$                        & \\
    $3^+$ &$\nu\frac{1}{2}^{-}[521]\otimes\nu\frac{5}{2}^{-}[512]$   &1.409 &                                       &\\
    $6^-$  &$\nu\frac{5}{2}^{-}[512]\otimes\nu\frac{7}{2}^{+}[633]$  &1.773  &$1.301^a$                      &$1.486$\\
    $5^-$  &$\nu\frac{1}{2}^{-}[521]\otimes\nu\frac{9}{2}^{+}[624]$  &2.124  & &\\
    $4^+$ &$\nu\frac{1}{2}^{-}[521]\otimes\nu\frac{7}{2}^{-}[514]$   &2.114  & &\\
    $5^+$ &$\nu\frac{5}{2}^{+}[523]\otimes\nu\frac{5}{2}^{+}[512]$  &2.448  &                                       &\\
\end{tabular}
\end{ruledtabular}
\end{table}
%%%%%%%%%%%%%%%%%%%%%%%%%%%%%%%%%%%%%%%%%%%%%

%%%%%%%%%%%%%%%%%%%%%%%%%%%%%%%%%%%%%%%%%%%%%
\begin{table}
\caption{\label{tab:Gd_K_isomer} Same as table~\ref{tab:Sm_K_isomer}, but for gadolinium isotopes. The experimental data are taken from Refs.~\cite{PatelZ2017_PRC96,PatelZ2014_PRL113}. }
\begin{ruledtabular}
\begin{tabular}{ccclcr}
    $K^\pi$&Configuration&$E_x$(MeV)& $E_x^{abc}$(MeV) & $E_{x}^{exp}$(MeV)\\ \hline
    $^{160}$Gd&&&&\\
    $4^+$  &$\pi\frac{5}{2}^{+}[413]\otimes\pi\frac{3}{2}^{+}[411]$    &0.813  & & \\
    $5^-$   &$\pi\frac{5}{2}^{+}[413]\otimes\pi\frac{5}{2}^{-}[532]$     &0.918  & &\\
    $6^-$   &$\pi\frac{5}{2}^{+}[413]\otimes\pi\frac{7}{2}^{-}[523]$     &1.425  &      &\\
    $5^-$   &$\pi\frac{3}{2}^{+}[411]\otimes\pi\frac{7}{2}^{-}[523]$     &1.739  &      &\\
    $4^+$  &$\nu\frac{3}{2}^{-}[521]\otimes\nu\frac{5}{2}^{-}[523]$    &1.492  &      &\\
    $5^-$   &$\nu\frac{5}{2}^{-}[523]\otimes\nu\frac{5}{2}^{+}[642]$   &2.004  && \\
    $^{162}$Gd&&&&\\
    $4^+$  &$\pi\frac{5}{2}^{+}[413]\otimes\pi\frac{3}{2}^{+}[411]$    &0.800  &   &\\
    $6^-$   &$\pi\frac{5}{2}^{+}[413]\otimes\pi\frac{7}{2}^{-}[523]$    &1.339  &   &\\
    $4^-$   &$\pi\frac{3}{2}^{+}[411]\otimes\pi\frac{5}{2}^{-}[532]$    &1.435  &         &\\
    $5^-$   &$\pi\frac{3}{2}^{+}[411]\otimes\pi\frac{7}{2}^{-}[523]$    &1.656  &           &\\
    $5^-$   &$\pi\frac{5}{2}^{+}[413]\otimes\pi\frac{5}{2}^{-}[532]$    &1.746  && \\
    $6^+$  &$\pi\frac{5}{2}^{-}[532]\otimes\pi\frac{7}{2}^{-}[523]$     &1.994  &  &\\
    $6^-$   &$\nu\frac{5}{2}^{-}[523]\otimes\nu\frac{7}{2}^{+}[633]$   &1.410 & &\\
    $3^+$  &$\nu\frac{1}{2}^{-}[521]\otimes\nu\frac{5}{2}^{-}[523]$   &1.700  &   &\\
    $5^-$   &$\nu\frac{3}{2}^{-}[521]\otimes\nu\frac{7}{2}^{+}[633]$   &1.944  &  &\\
    $^{164}$Gd&&&&\\
    $4^+$  &$\pi\frac{5}{2}^{+}[413]\otimes\pi\frac{3}{2}^{+}[411]$   &0.823  &  & \\
    $6^-$   &$\pi\frac{5}{2}^{+}[413]\otimes\pi\frac{7}{2}^{-}[523]$   &1.315  &   & \\
    $4^-$   &$\pi\frac{3}{2}^{+}[411]\otimes\pi\frac{5}{2}^{-}[532]$   &1.517  &    & \\
    $5^-$   &$\pi\frac{3}{2}^{+}[411]\otimes\pi\frac{7}{2}^{-}[523]$   &1.631  &    & \\
    $5^-$   &$\pi\frac{5}{2}^{+}[413]\otimes\pi\frac{5}{2}^{-}[532]$   &1.844  &  &\\
    $6^+$  &$\pi\frac{5}{2}^{-}[532]\otimes\pi\frac{7}{2}^{-}[523]$   &1.851  &    & \\
    $4^-$   &$\nu\frac{1}{2}^{-}[521]\otimes\nu\frac{7}{2}^{+}[633]$   &1.168 & &1.096\\
    $3^+$   &$\nu\frac{5}{2}^{-}[523]\otimes\nu\frac{1}{2}^{-}[521]$   &1.875  &&\\
    $6^-$   &$\nu\frac{5}{2}^{-}[512]\otimes\nu\frac{7}{2}^{+}[633]$   &1.887      &      &\\
    $3^+$   &$\nu\frac{1}{2}^{-}[521]\otimes\nu\frac{5}{2}^{-}[512]$   &2.166          &  &\\
    $^{166}$Gd& &&&\\
    $4^+$  &$\pi\frac{5}{2}^{+}[413]\otimes\pi\frac{3}{2}^{+}[411]$   &0.839  &$1.300^a$ &  1.350 \\
    $6^-$   &$\pi\frac{5}{2}^{+}[413]\otimes\pi\frac{7}{2}^{-}[523]$   &1.301  &                  & \\
    $4^-$   &$\pi\frac{3}{2}^{+}[411]\otimes\pi\frac{5}{2}^{-}[532]$   &1.506  &$1.769^a$  &      \\
    $5^-$   &$\pi\frac{3}{2}^{+}[411]\otimes\pi\frac{7}{2}^{-}[523]$   &1.609  &                  &     \\
    $6^+$  &$\pi\frac{5}{2}^{-}[532]\otimes\pi\frac{7}{2}^{-}[523]$   &1.753   &                  &  \\
    $5^-$   &$\pi\frac{5}{2}^{+}[413]\otimes\pi\frac{5}{2}^{-}[532]$   &1.888  &$1.826^a$  & \\
    $3^+$  &$\nu\frac{1}{2}^{-}[521]\otimes\nu\frac{5}{2}^{-}[512]$  &1.316  &$1.400^a$  &  \\
    $6^-$   &$\nu\frac{5}{2}^{-}[512]\otimes\nu\frac{7}{2}^{+}[633]$  &1.690  &$1.288^a$  &    1.601   \\
    $5^-$   &$\nu\frac{1}{2}^{-}[521]\otimes\nu\frac{9}{2}^{+}[624]$  &1.981  & &\\
    $4^+$  &$\nu\frac{7}{2}^{-}[514]\otimes\nu\frac{1}{2}^{-}[521]$   &2.031  & &\\
    $8^+$  &$\nu\frac{7}{2}^{+}[633]\otimes\nu\frac{9}{2}^{+}[624]$  &2.345  &&\\
    $7^-$   &$\nu\frac{7}{2}^{+}[633]\otimes\nu\frac{7}{2}^{-}[514]$   &2.388   && \\
    $5^+$  &$\nu\frac{5}{2}^{-}[523]\otimes\nu\frac{5}{2}^{-}[512]$    &2.403   && \\
    $4^-$   &$\nu\frac{7}{2}^{+}[633]\otimes\nu\frac{1}{2}^{-}[521]$   &2.494   &$1.684^a$&\\
\end{tabular}
\end{ruledtabular}
\end{table}
%%%%%%%%%%%%%%%%%%%%%%%%%%%%%%%%%%%%%%%%%%%%%

%%%%%%%%%%%%%%%%%%%%%%%%%%%%%%%%%%%%%%%%%%%%%
\begin{figure}%[!]
\includegraphics[scale=0.5]{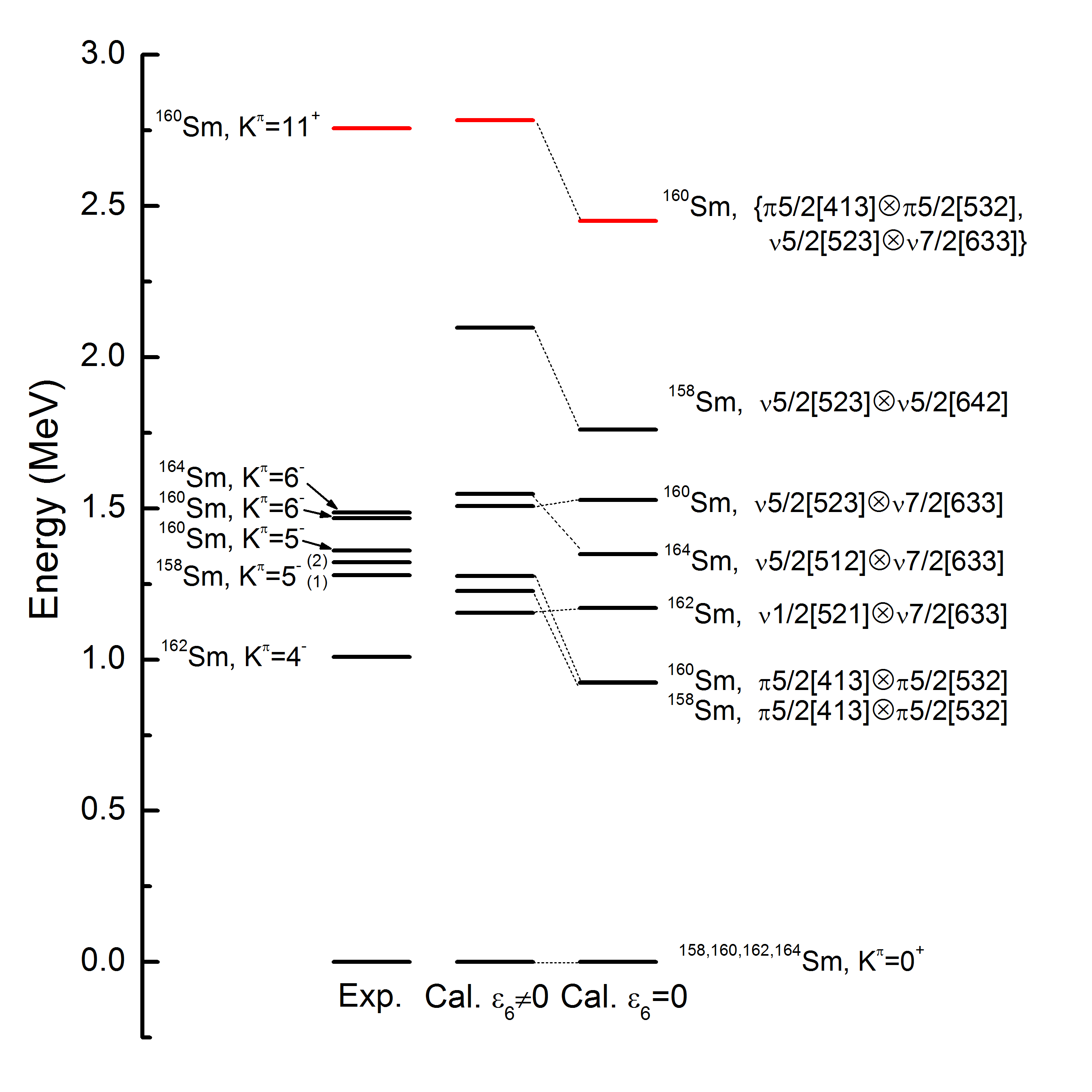}
\caption{\label{fig:BandHeadE_Sm}
     (Color online) The experimental (left) and calculated band head energy with $\varepsilon_6\neq0$ (middle) and $\varepsilon_6=0$ (right) for samarium isotopes. The experimental data are taken from Refs.~\cite{PatelZ2017_PRC96,PatelZ2016_PLB753,PatelZ2014_PRL113,WangE2014_PRC90,SimpsonG2009_PRC80,nndc}. $K^\pi=5^-(1)$ and $K^\pi=5^-(2)$ in \element{158}Sm denote the states taken from Simpson \textit{et al.} in Ref.~\cite{SimpsonG2009_PRC80} and from Wang \textit{et al.} in Ref~\cite{WangE2014_PRC90}, respectively. The positive-parity (negative-parity) levels are denoted by red (black) lines.}
\end{figure}
%%%%%%%%%%%%%%%%%%%%%%%%%%%%%%%%%%%%%%%%%%%%%

%%%%%%%%%%%%%%%%%%%%%%%%%%%%%%%%%%%%%%%%%%%%%
\begin{figure}%[!]
\includegraphics[scale=0.5]{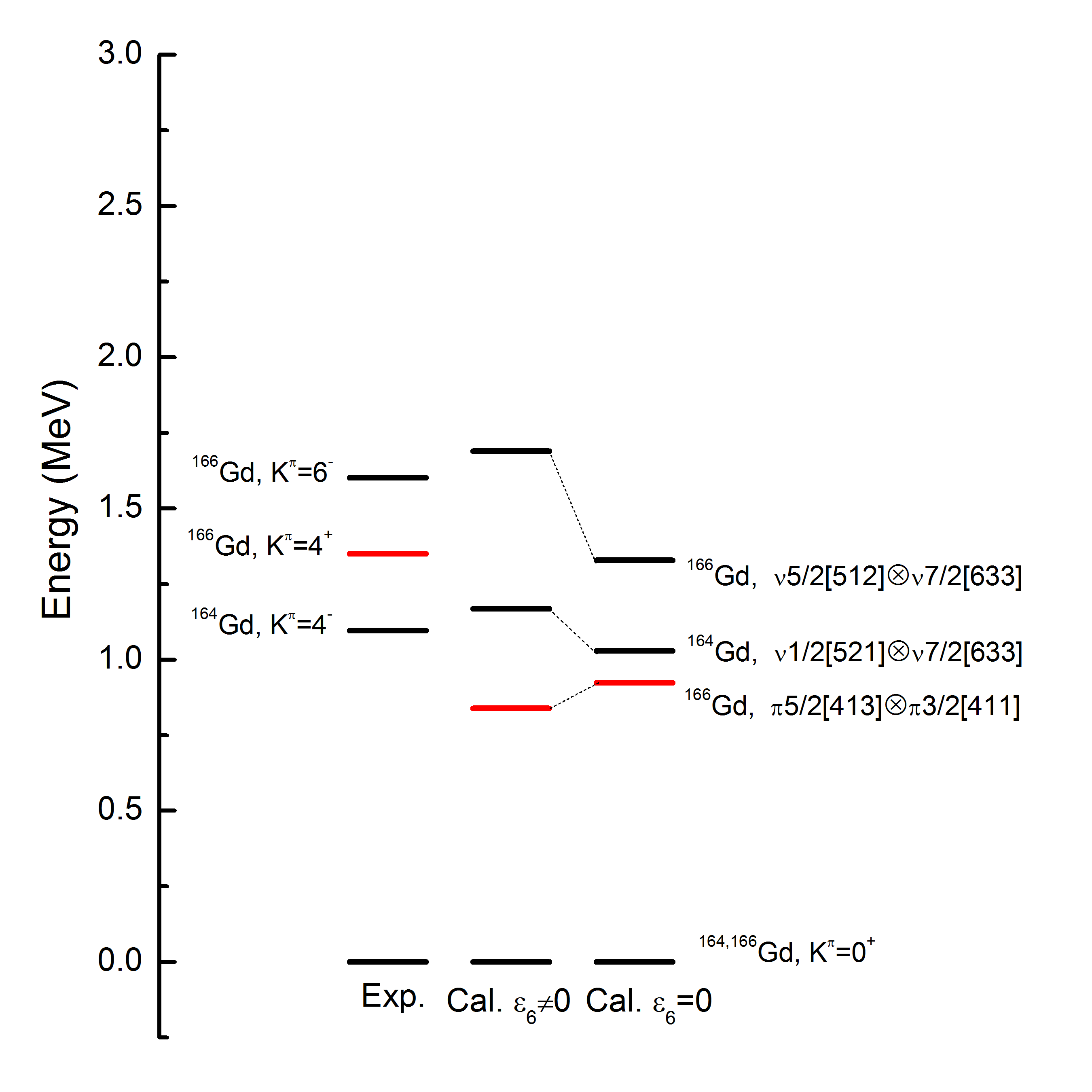}
\caption{\label{fig:BandHeadE_Gd}
     (Color online) Same as Fig.~\ref{fig:BandHeadE_Sm}, but for gadolinium isotopes. The experimental data are taken from Refs.~\cite{PatelZ2017_PRC96,PatelZ2014_PRL113}.}
\end{figure}
%%%%%%%%%%%%%%%%%%%%%%%%%%%%%%%%%%%%%%%%%%%%%

The low-lying multi-particle states of samarium and gadolinium isotopes predicted by the PNC-CSM method are listed in Table~\ref{tab:Sm_K_isomer} and~\ref{tab:Gd_K_isomer}, respectively.  Among these, comparison with the available experimental data are displayed in Fig.~\ref{fig:BandHeadE_Sm} and~\ref{fig:BandHeadE_Gd}, respectively. The experimental data are taken from Refs.~\cite{PatelZ2017_PRC96,PatelZ2016_PLB753,PatelZ2014_PRL113,WangE2014_PRC90,SimpsonG2009_PRC80,PatelZ2016_EWoC123,nndc}. The results with $\varepsilon_{6}=0$ are displayed to examine the high-order deformation effect. In general, compared to the $\varepsilon_{6}=0$ calculations, the non-zero $\varepsilon_{6}$ results reproduce the experimental multi-particle state energies better, except for the $1.486$ MeV $6^-$ state in $^{164}$Sm and the $1.350$ MeV $4^+$ state in $^{166}$Gd~\cite{PatelZ2014_PRL113}. 

Note that the energy $1.486$ MeV of the 0.60(0.14) $\mu$s $6^-$ isomer in $^{164}$Sm is obtained by assuming energy of the first $2^+$ level at 0.069 MeV from rotational band systematics of nuclei in this mass region~\cite{nndc}. Suggested by potential energy surface calculations, the two-neutron $\nu\frac{5}{2}^{-}[512]\otimes\nu\frac{7}{2}^{+}[633]$ configuration was assigned to the $6^-$ isomer~\cite{PatelZ2014_PRL113}, which is confirmed in the present PNC-CSM configuration assignment. According to the tables of M\"oller \textit{et al.}, the $\varepsilon_{6}$ (=0.053) maximizes for $^{164}$Sm in samarium isotopes. With such large $\varepsilon_{6}$ value, the state energy of $6^-$ is shifted up by 579 keV (compared to $\varepsilon=0$ calculations), which leads to a worse reproduction of the experimental data. The deformation parameters are recalculated and the value of $\varepsilon_{6}$ equals $0.04$ in the new table of M\"oller \textit{et al.} in 2016~\cite{MoellerP2016_ADNDT109-110}. When $\varepsilon_{6}=0.04$ is adopted, a lower/better state energy of $1.773$ MeV can be obtained in the PNC-CSM calculation. Potential energy surface calculation suggests an even smaller value of $\beta_{6}=-0.02$. By using this value, 1.301 MeV state energy is obtained by the potential energy surface calculations. The significant $\varepsilon_{6}$ influence on the $6^-$ isomer in $^{164}$Sm originates from the $\varepsilon_{6}$ effects on the single-particle levels in Fig.~\ref{fig:Nilsson}.  By including the non-zero $\varepsilon_{6}$, the energy space between the $\nu\frac{7}{2}^{+}[633]$ and $\nu\frac{5}{2}^{-}[512]$ orbitals enlarges and the deformed energy gap at $N=102$ arises, which results in a higher involved multi-particle state energy. Considering that other multi-particle states involved with the $\nu\frac{7}{2}^{+}[633]$ orbital, like two-neutron $6^-$ state in $^{160}$Sm and $4^-$ state in $^{162}$Sm, reproduce the experimental data well, the disagreement of the two-neutron $6^-$ state in $^{164}$Sm is mainly caused by the upward shift of the $\nu\frac{5}{2}^{-}[512]$ orbital with non-zero $\varepsilon_{6}$. This indicates that the energy gap at $N=102$ should be smaller. 

Significant $\varepsilon_{6}$ effect is also demonstrated on the multi-proton states. Take the $\pi\frac{5}{2}^{+}[413]\otimes\pi\frac{5}{2}^{-}[532]$ state for example, compared to $\varepsilon_{6}=0$ calculations, the energies of the $\pi\frac{5}{2}^{+}[413]\otimes\pi\frac{5}{2}^{-}[532]$ states in $^{158,160}$Sm are increased by $296-359$ keV, which gives a better reproduction of the experiment data. As a consequence, the four-particle state $11^{+}$ in $^{160}$Sm is lifted up, and agrees better with the experiment data. The energies of the $\pi\frac{5}{2}^{+}[413]\otimes\pi\frac{5}{2}^{-}[532]$ states are mainly gained from the enlarged $Z=62$ energy gap of the single proton Nilsson levels with non-zero $\varepsilon_{6}$ (see Fig.~\ref{fig:Nilsson}). Here is an evidence that the proton $Z=62$ energy gap seems necessary to reproduce well the experimental multi-particle states. 

The two-particle $5^-$ side band of \element{158}{Sm}, which built on top of the $1.279$ MeV isomeric state, was firstly identified by Zhu \textit{et al.} and a two-neutron $\nu\frac{5}{2}^{-}[523]\otimes\nu\frac{5}{2}^{+}[624]$ configuration was assigned~\cite{ZhuS1995_JoPGNaPP21}. Simpson \textit{et al.} extended the band to higher spin up to $I=18$ in Ref.~\cite{SimpsonG2009_PRC80}. For convenience, this state is denoted as $K^{\pi}=5^- (1)$ hereinafter. Wang \textit{et al.} reinvestigated the high-spin states of \element{158}{Sm}~\cite{WangE2014_PRC90}. A new two-particle $5^-$ side band was observed and a two-proton $\pi\frac{5}{2}^{+}[413]\otimes\pi\frac{5}{2}^{-}[532]$ configuration was assigned to the $1.322$ MeV band head. To distinguish this newly observed $5^-$ band from the earlier one, the $1.322$ MeV $5^-$ state is denoted as $K^{\pi}=5^- (2)$ band hereinafter.  

The present PNC-CSM calculations predict three $5^-$ states in \element{158}{Sm} (see Table.~\ref{tab:Sm_K_isomer}). One of them is the two-proton 1.227 MeV state with the configuration of $\pi\frac{5}{2}^{+}[413]\otimes\pi\frac{5}{2}^{-}[532]$, which is the lowest 2-particle state in \element{158}{Sm}. The other two are neutron states with much higher energies, i. e. the 2.097 MeV and 2.165 MeV states with configurations of $\nu\frac{5}{2}^{-}[523]\otimes\nu\frac{5}{2}^{+}[624]$ and $\nu\frac{3}{2}^{-}[521]\otimes\nu\frac{7}{2}^{+}[633]$, respectively. Analysed together with the rotational bands on $5^-$ states (see below for moment of inertia), the present PNC-CSM calculations based on the Nilsson single-particle levels suggest that the $1.279$ MeV $5^- (1)$ state is the two-proton $\pi\frac{5}{2}^{+}[413]\otimes\pi\frac{5}{2}^{-}[532]$ configuration state. The next $5^-$ state by the PNC-CSM calculation is the 2.097 MeV two-neutron $\nu\frac{5}{2}^{-}[523]\otimes\nu\frac{5}{2}^{+}[624]$ configuration state. The observed $1.322$ MeV $5^- (2)$ state is tentatively assigned as the two-neutron $\nu\frac{5}{2}^{-}[523]\otimes\nu\frac{5}{2}^{+}[624]$ configuration state but keep in mind that both the calculated state energy and moment of inertia are larger than the experimental data. 

The low-lying two-particle $5^-$ isomeric state (with energy of 1.361 MeV) was also observed in \element{160}{Sm}~\cite{SimpsonG2009_PRC80,PatelZ2016_PLB753}. It was previously suggested to be a two-neutron $\nu\frac{5}{2}^{-}[523]\otimes\nu\frac{5}{2}^{+}[624]$ configuration state for its similarity to the $5^-$ state in the neighboring samarium isotopes~\cite{SimpsonG2009_PRC80}. However, blocked-BCS calculations assigned it as a two-proton state of $\pi\frac{5}{2}^{+}[413]\otimes\pi\frac{5}{2}^{-}[532]$ configuration recently~\cite{PatelZ2016_PLB753}, which is confirmed by the present PNC-CSM calculation. In addition to the two-particle $5^-$ isomeric state, a new band structure on top of the $6^-$ state and a four-particle $11^+$ isomeric state are recognized in \element{160}{Sm} recently~\cite{PatelZ2016_PLB753}. The PNC-CSM calculations show that the two-proton $5^-$ state with configuration $\pi\frac{5}{2}^{+}[413]\otimes\pi\frac{5}{2}^{-}[532]$ and the two-neutron $6^-$ state with configuration $\nu\frac{5}{2}^{-}[523]\otimes\nu\frac{7}{2}^{+}[633]$ are the lowest two-proton and neutron state, respectively. The combination of these two states forms the lowest four-particle 2.784 MeV $11^+$ state which reproduce the experiment date (2.757 MeV) quite well. 

Two-proton $4^+$ state with $\pi\frac{5}{2}^{+}[413]\otimes\pi\frac{3}{2}^{+}[411]$ configuration is predicted to be the lowest proton state in the gadolinium isotopes. Experimentally, $4^+$ state was identified at 1.350 MeV in $^{166}$Gd by Patel \textit{et al.}, to which the $\pi\frac{5}{2}^{+}[413]\otimes\pi\frac{3}{2}^{+}[411]$ configuration was suggested by the potential energy surface calculation~\cite{PatelZ2014_PRL113}. The present PNC-CSM calculation gives a too low energy at 0.839 MeV. It can be seen in Fig.~\ref{fig:BandHeadE_Gd} that, compared with the result of zero $\varepsilon_{6}$ calculation (0.923 MeV), result with non-zero $\varepsilon_{6}$ agrees even worse with the experimental data. The underestimation of the $4^+$ state energy by the PNC-CSM calculation indicates that the $\pi\frac{5}{2}^{+}[413]$ and $\pi\frac{3}{2}^{+}[411]$ orbitals at $Z=64$ (see Fig.~\ref{fig:Nilsson}) locate too close to each other. 

The experimental observed $950(60)$ ns $6^-$ isomeric state in $^{166}$Gd is assigned as the two-neutron $\nu\frac{5}{2}^{-}[512]\otimes\nu\frac{7}{2}^{+}[633]$ configuration state which is consistent with the assignment of the potential energy surface calculation~\cite{PatelZ2014_PRL113}. However, the lowest two-neutron state in the $N=102$ isotone \element{166}{Gd} and \element{164}{Sm} is predicted as the $3^+$ state with configuration $\nu\frac{1}{2}^{-}[521]\otimes\nu\frac{5}{2}^{-}[512]$ in the PNC-CSM calculations. 

The $4^-$ isomeric state in $N=100$ isotones $^{164}$Gd and $^{162}$Sm was discovered by Yokoyama \textit{et al.} in Ref.~\cite{YokoyamaR2017_PRC95}. Deformed Hartree-Fock and projected shell model calculations interpreted it as the two-neutron $\nu\frac{1}{2}^{-}[521]\otimes\nu\frac{7}{2}^{+}[633]$ configuration, which is as same as known $4^-$ isomers in $N=100$ isotones $^{168}$Er and $^{170}$Yb. Soon after, new data was identified independently by Patel \textit{et al.}, and the configuration assignment was confirmed by the Nilsson-BCS calculations~\cite{PatelZ2014_PRL113}. The present PNC-CSM calculations can reproduce the experimental $4^-$ isomer very well and confirm the two-neutron $\nu\frac{1}{2}^{-}[521]\otimes\nu\frac{7}{2}^{+}[633]$ configuration assignment. 

\subsection{Occupation probability}

%%%%%%%%%%%%%%%%%%%%%%%%%%%%%%%%%%%%%%%%%%%%%
\begin{figure}
\includegraphics[scale=0.8]{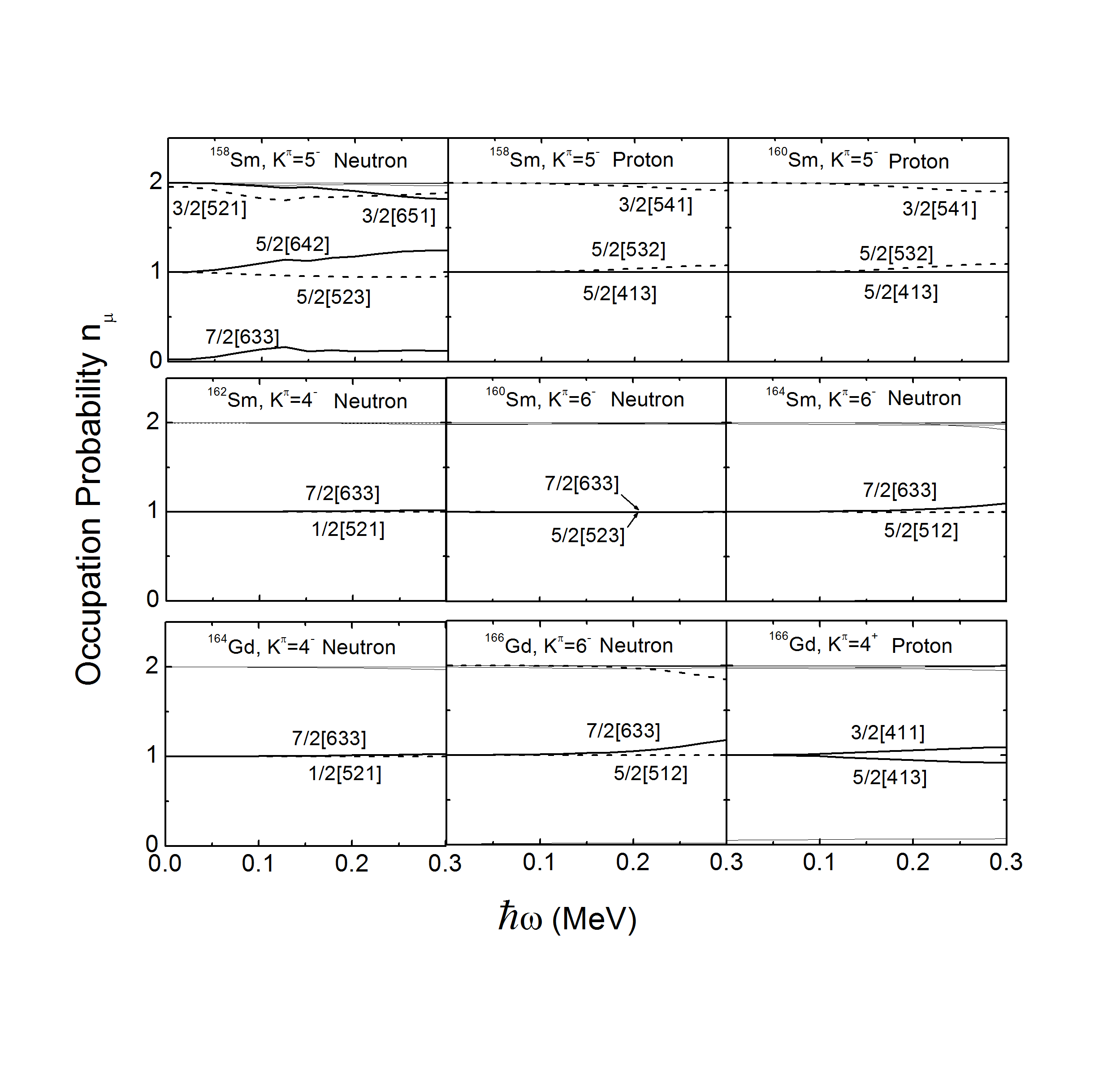}
\caption{\label{fig:Pr}
Occupation probabilities $n_\mu$ of cranked Nilsson orbital $\mu$ (including both $\alpha = \pm1/2$) near the Fermi surface of the samarium and gadolinium isotopes for the two-particle states bands. The thick solid (dashed) lines denote positive (negative) parity orbitals. Fully occupied $n_{\mu}\approx2$ and empty $n_{\mu}\approx0$ orbitals denoted by thin lines are not labelled.}
\end{figure}
%%%%%%%%%%%%%%%%%%%%%%%%%%%%%%%%%%%%%%%%%%%%%

The configuration of each multi-particle state is explicitly determined through the occupation probability $n_{\mu}$ of each cranked Nilsson orbital $\mu$. Once the wave function (Eq.~\ref{eq:eigenstate}) is obtained, the occupation probability of an orbital $\mu$ can be calculated as,
\begin{equation}
\label{eq:n_mu}
    n_{\mu} = \sum_{i}|C_{i}|^{2}P_{i\mu},
\end{equation} 
where $P_{i\mu}=1$ if $|\mu\rangle$ is occupied and $P_{i\mu}=0$ otherwise. The total particle number $N=\sum_\mu n_\mu$. The rotational frequency $\omega$-dependence of occupation probabilities $n_\mu$ can also give more detailed informations of rotational properties, like band-crossing, configuration mixing and so on. 

In Fig.~\ref{fig:Pr} it shows the occupation probabilities $n_{\mu}$ versus frequency $\hbar\omega$ of each cranked Nilsson orbital $\mu$ near the Fermi surface of samarium and gadolinium isotopes, where $\mu$ includes both of $\alpha=\pm\frac{1}{2}$. $|\mu\rangle$ is blocked at $n_{\mu}\approx 1$ while it is fully occupied and empty at $n_{\mu}\approx 2$ and $n_{\mu}\approx 0$, respectively. We have checked that $n_{\mu}$ for ground state bands (GSB) displays no band-crossing, and the configuration assignment is not necessary for GSB in the even-even nuclei. Therefore only the $n_{\mu}$ of the experimental observed two-particle states are presented. It is seen that the configurations for these two-particle states are quite pure, especially at the low frequency region, except for the two-neutron $K^\pi=5^-$ band in $^{158}$Sm. 

For the two-neutron $5^-$ band in $^{158}$Sm, the blocked neutron orbitals are $\nu\frac{5}{2}[642]$ and $\nu\frac{5}{2}[523]$. Due to the configuration mixing, $n_{\mu}$ of $\nu\frac{5}{2}[642]$ orbital increases with $n_{\mu}>1$ at frequency $\hbar\omega>0.05$ MeV. Meanwhile, $n_{\mu}$ of the $\nu\frac{7}{2}[633]$ orbital above the Fermi surface increases and it of the $\nu\frac{3}{2}[651]$ and $\nu\frac{3}{2}[521]$ orbitals below the Fermi surface decreases. This leads to the decreasing trend of the moment of inertia with rotational frequency shown in Fig.~\ref{fig:MOI}. 

\subsection{Moment of inertia}

%%%%%%%%%%%%%%%%%%%%%%%%%%%%%%%%%%%%%%%%%%%%%
\begin{figure}%[!]
\includegraphics[scale=0.55]{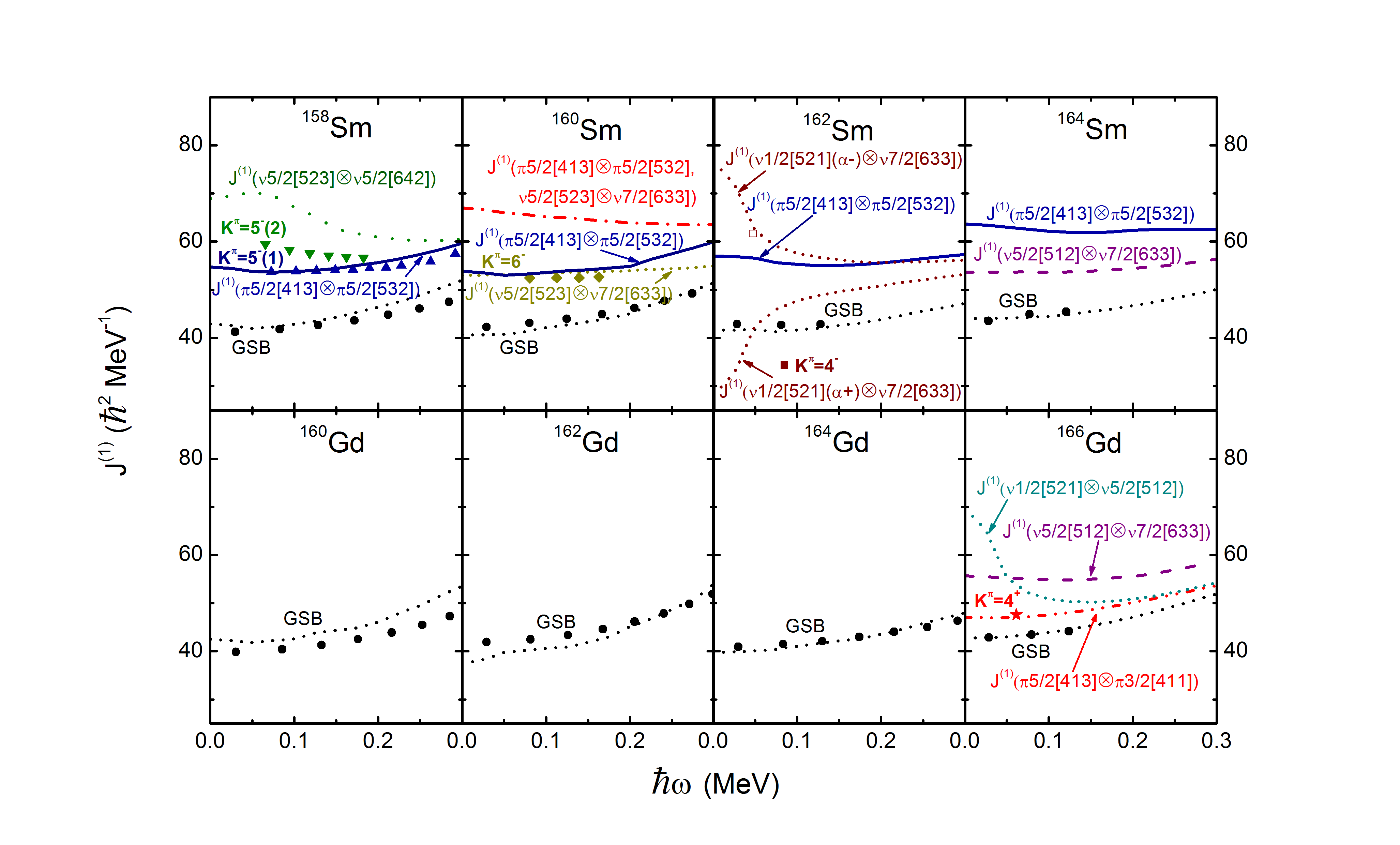}
\caption{\label{fig:MOI}
(color online) Calculated kinematic moments of inertia for samarium and gadolinium isotopes, compared with available experimental data~\cite{SimpsonG2009_PRC80,WangE2014_PRC90,PatelZ2014_PRL113,PatelZ2016_PLB753,PatelZ2016_EWoC123,nndc}. Experimental data are denoted by symbols and the multi-particle state configurations of these bands are labelled by $K^\pi$. Data for the $K^\pi=5^-(1)$ and $K^\pi=5^-(2)$ bands in \element{158}Sm are taken from Simpson \textit{et al.} in Ref.~\cite{SimpsonG2009_PRC80} and from Wang \textit{et al.} in Ref~\cite{WangE2014_PRC90}, respectively. Theoretical calculations are denoted by lines and their configurations are labelled as $J^{(1)}(\mu\otimes\nu)$. }
\end{figure}
%%%%%%%%%%%%%%%%%%%%%%%%%%%%%%%%%%%%%%%%%%%%%

The angular momentum alignment $\left\langle J_{x} \right\rangle$ of
the state $\left\vert \psi \right\rangle$ is given by
\begin{equation}
 \left\langle \psi \right| J_{x}
 \left| \psi \right\rangle
 = \sum_{i}\left|C_{i}\right| ^{2}
   \left\langle i\right| J_{x}\left| i\right\rangle
 + 2\sum_{i<j}C_{i}^{\ast }C_{j}
   \left\langle i\right| J_{x}\left| j\right\rangle\ .
 \label{eq:Jx}
\end{equation}
The kinematic moment of inertia is $\mathcal{J}^{(1)}=\left\langle \psi \right\vert J_{x}\left\vert\psi \right\rangle /\omega $. In Fig.~\ref{fig:MOI} it shows the comparison of the calculated kinematic moments of inertia and the available experimental data~\cite{SimpsonG2009_PRC80, WangE2014_PRC90, PatelZ2014_PRL113, PatelZ2016_PLB753, PatelZ2016_EWoC123, nndc} for samarium and gadolinium isotopes. The experimental data are reproduced well by the theoretical results. According to the PNC-CSM calculations, only the neutron $4^-$ band of \element{162}{Sm} and $3^+$ band of \element{166}{Gd} show an obvious signature splitting. Both bands are concerned with the $\nu\frac{1}{2}[521]$ orbital. There is no signature splitting occurring for other bands. Thus, unless it is labelled explicitly, only the calculated favored ($\alpha$=0) signature bands are presented in Fig.~\ref{fig:MOI}. As it is shown by the occupation probability in Fig.~\ref{fig:Pr}, there is no band-crossing involved in these multi-particle state bands, neither is in the GSB. Therefore all the moments of inertia display a gradual and smooth variation with frequency $\hbar\omega$.    

For the $K^\pi=5^-$ state bands of $^{158}$Sm, the calculated moment of inertia of two-proton $\pi\frac{5}{2}^{+}[413]\otimes\pi\frac{5}{2}^{-}[532]$ state are in good agreement with the experimental 1.279 MeV $5^-$ state band of Simpson \textit{et al.}~\cite{SimpsonG2009_PRC80}. This is consistent with the configuration assignment by band head energy stated above. The calculated moments of inertia for the two-neutron $\nu\frac{5}{2}^{-}[523]\otimes\nu\frac{5}{2}^{+}[642]$ state are larger than the experimental 1.322 MeV $5^-$ state band for the whole observed rotational frequency though the decreasing trend can be displayed.  

The 1.7$\mu$s  $4^-$ isomer was identified in $^{162}$Sm by Patel \textit{et al.}~\cite{PatelZ2016_EWoC123,PatelZ2017_PRC96} and Yokoyama \textit{et al.}~\cite{YokoyamaR2017_PRC95} independently. The configuration is assigned as a two-neutron $\nu\frac{1}{2}^{-}[521]\otimes\nu\frac{7}{2}^{+}[633]$ configuration by Nilsson-BCS~\cite{PatelZ2016_EWoC123,PatelZ2017_PRC96}, deformed Hartree-Fock~\cite{YokoyamaR2017_PRC95}, projected shell model~\cite{YangY2010_JoPGNaPP37} and present PNC-CSM calculations. In Ref.~\cite{PatelZ2016_EWoC123}, a $146$ keV $\gamma$ ray is visible, which is tentatively placed as a transition from a $5^-$ state to the isomeric $4^-$ state. According to this assignment, moment of inertia is extracted as the solid square shown in Fig.~\ref{fig:MOI}, which is even lower than the ground state band and the PNC-CSM calculation can not reproduce it well. Normally, due to the Coriolis effect and the pairing reduction, moment of inertia of the multi-particle state should be larger than the ground state. The moment of inertia is reextracted by assuming that the $146$ keV $\gamma$ ray is the transition from a $6^-$ state to the isomeric $4^-$ state, which is shown by the hollow square in Fig.~\ref{fig:MOI}. Then the theoretical calculation gives better agreement of the experimental data. Therefore, the present calculation prefers the $146$ keV $\gamma$ ray to be the transition from a $6^-$ state.

\subsection{Electromagnetic property}

%%%%%%%%%%%%%%%%%%%%%%%%%%%%%%%%%%%%%%%%%%%%%%
%\begin{figure}
%\includegraphics[scale=0.4]{BE2_Sm}
%\caption{\label{fig:BE2}
%    The predicted $B(E2)$ of GSB and proton two-qp isomeric $K^\pi=5^-$ bands in samarium isotopes by PNC-CSM calculation. The thickness of dashed lines are denoted the $\omega$-dependence of $B(E2)$, and the thickest denotes $\hbar\omega=0$. The insert is established to see the $\omega$-dependent $B(E2)$ of GSBs clearly. }
%\end{figure}
%%%%%%%%%%%%%%%%%%%%%%%%%%%%%%%%%%%%%%%%%%%%%

%%%%%%%%%%%%%%%%%%%%%%%%%%%%%%%%%%%%%%%%%%%%%
\begin{figure}
\includegraphics[scale=0.4]{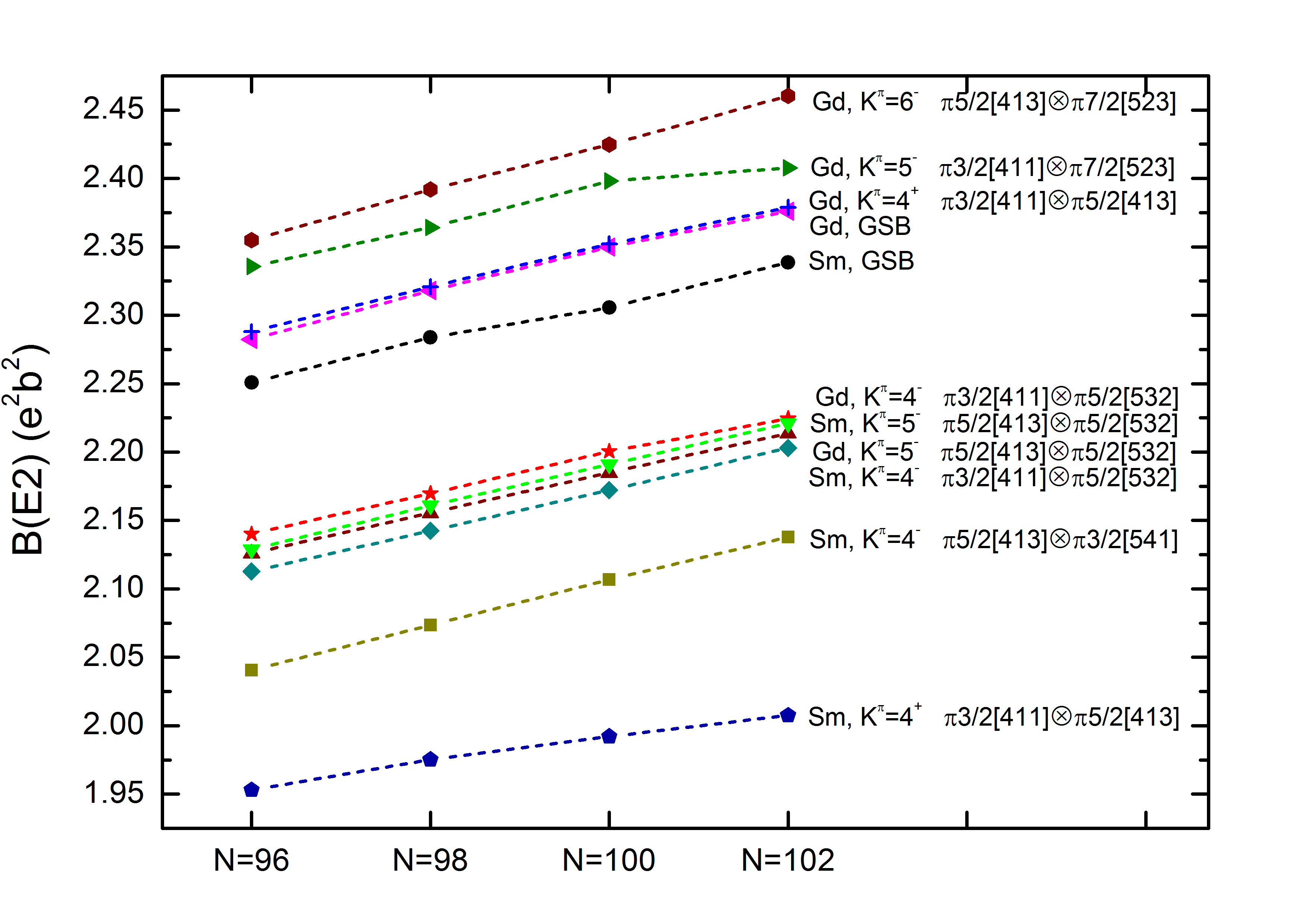}
\caption{\label{fig:BE2_isotope}
(color online) Calculated $B(E2,\omega=0)$ values systematics, connected by the dashed lines, of the GSB and two-particle states for samarium and gadolinium isotopes.}
\end{figure}
%%%%%%%%%%%%%%%%%%%%%%%%%%%%%%%%%%%%%%%%%%%%%

%%%%%%%%%%%%%%%%%%%%%%%%%%%%%%%%%%%%%%%%%%%%%
\begin{figure}
\includegraphics[scale=0.4]{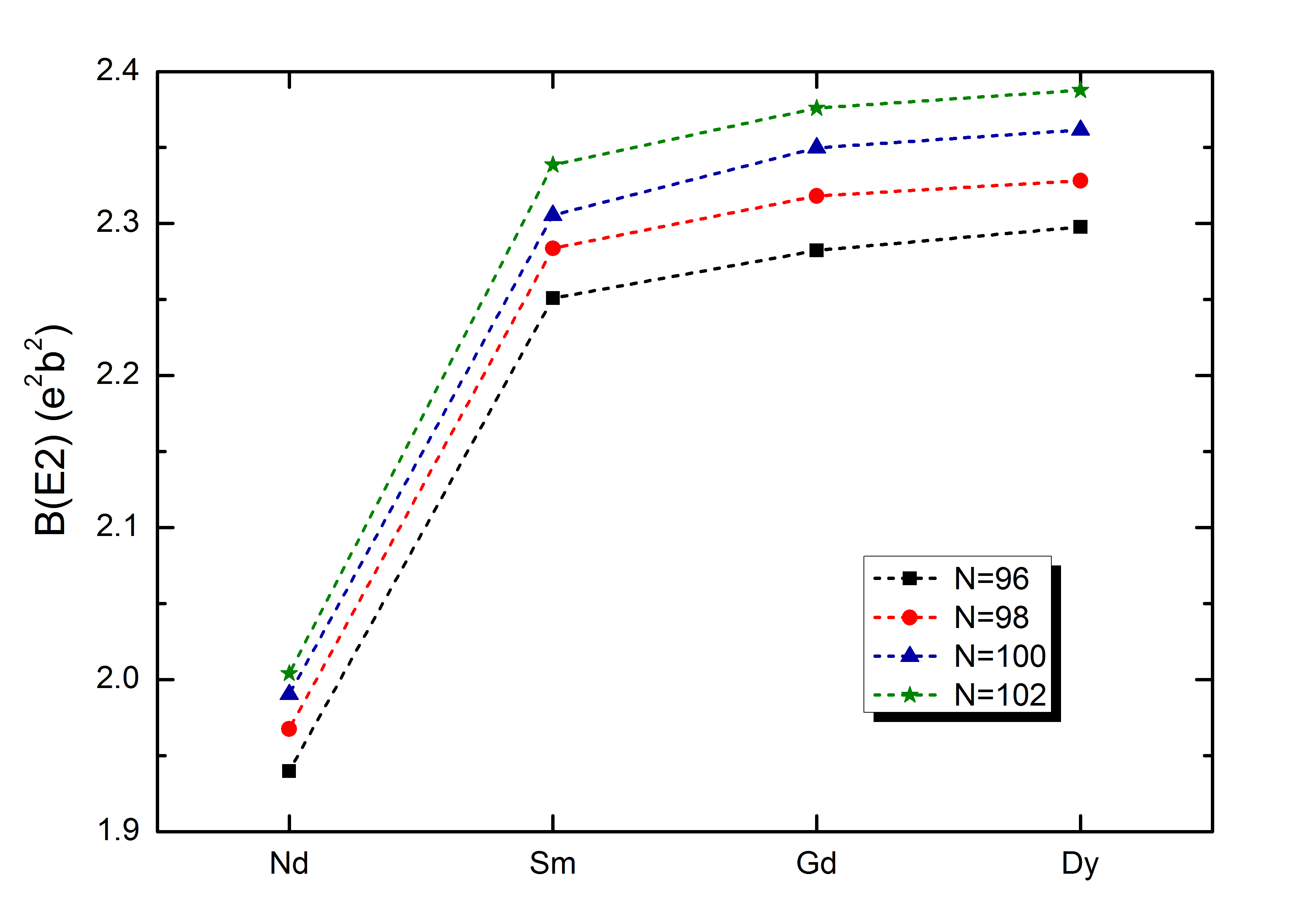}
\caption{\label{fig:BE2_isotone}
(color online) Same as Fig.~\ref{fig:BE2_isotope}, but for the GSB of $N=96,98,100,102$ isotones.}
\end{figure}
%%%%%%%%%%%%%%%%%%%%%%%%%%%%%%%%%%%%%%%%%%%%%

%%%%%%%%%%%%%%%%%%%%%%%%%%%%%%%%%%%%%%%%%%%%%
\begin{figure}
\includegraphics[scale=0.5]{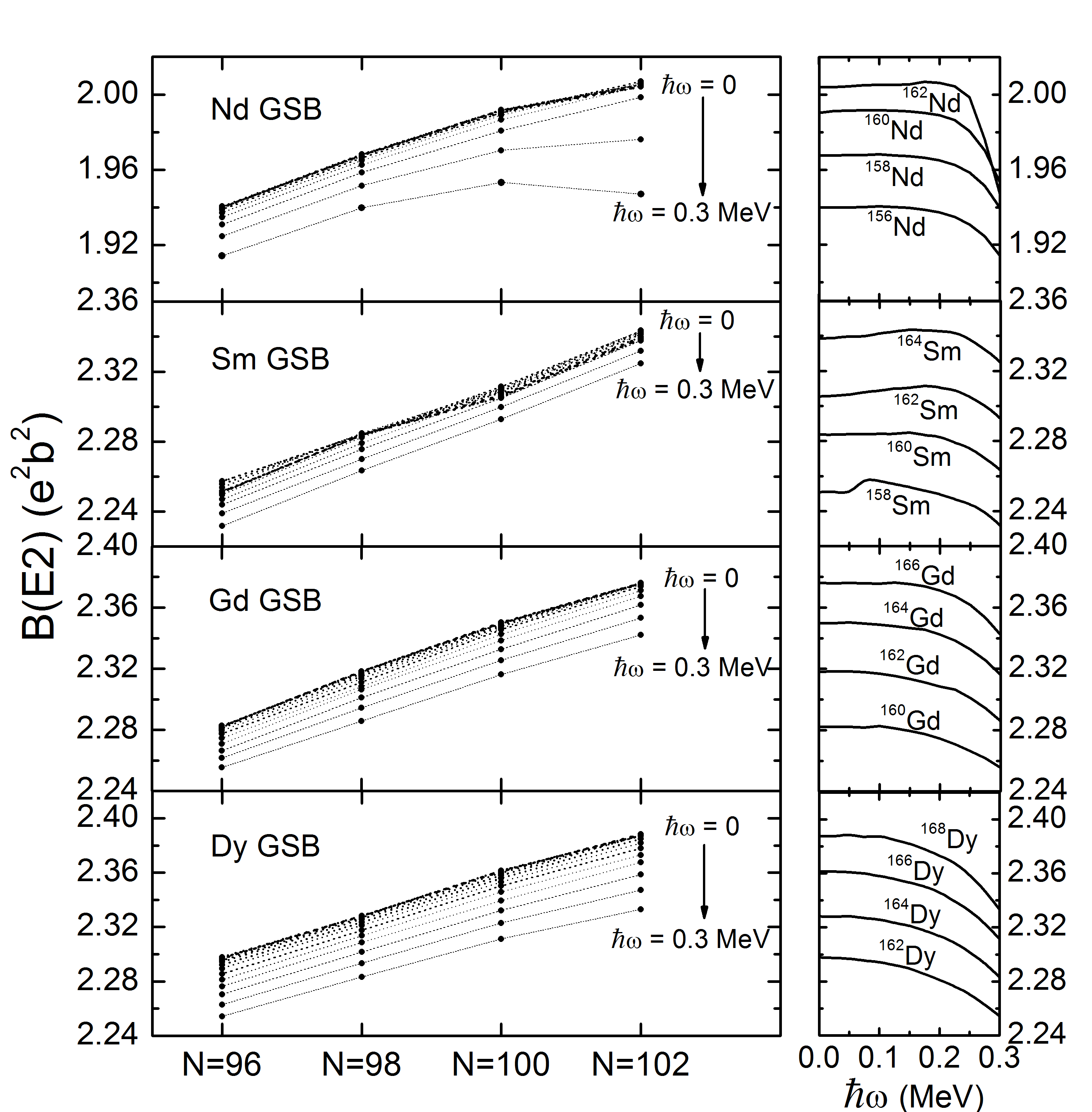}
\caption{\label{fig:BE2}
(left panel) Calculated $\omega$-dependent $B(E2)$ values systematics, connected by the dashed lines, of the GSB for neodymium, samarium, gadolinium and dysprosium isotopes. The thickness of the dashed lines denotes the $\omega$-dependence of $B(E2)$. (right panel) $B(E2)$ values versus rotational frequency $\hbar\omega$ of the GSB for neodymium, samarium, gadolinium and dysprosium isotopes.}
\end{figure}
%%%%%%%%%%%%%%%%%%%%%%%%%%%%%%%%%%%%%%%%%%%%%

The electromagnetic transition is useful to test the nuclear wave function and to deduce the nuclear collectivity informations. With eigenstate $|\psi\rangle$ of the cranked shell model Hamiltonian is obtained, the electronic quadrupole transition probabilities $B(E2)$ can be derived in the semiclassical approximation as,
\begin{equation}
    B(E2) = \frac{3}{8} \langle \psi | Q_{20}^p | \psi \rangle^2,
    \label{eq:BE2}
\end{equation}
where $Q_{20}^p$ corresponds to the laboratory quadrupole moments of protons,
\begin{equation}
    Q_{20} = r^2 Y_{20} = \sqrt{\frac{5}{16\pi}}(3z^2-r^2).
\end{equation}

Since the valence single-particle space is constructed in the major shells from N = 0 to N = 5 (N=6) for proton (neutron), there is no effective charge involved. $B(E2)$ value in Eq.~\ref{eq:BE2} is extremely sensitive to the quadruple deformation parameters which can not be obtained self-consistently in the PNC-CSM method. To avoid the effect from the parameters,  $\varepsilon_{2}=0.275$ is used to calculate the $B(E2)$ values for all the nuclei considered in \textit{this section}. Therefore, the systematic behavior of the $B(E2)$ values along an isotone or isotopic chain is a pure microscopic effect of the nuclear many-body wave functions.    

In Fig.~\ref{fig:BE2_isotope}, it demonstrates the $B(E2,\omega=0)$ value (in $e^2b^2$) systematics of the GSB and two-particle states for samarium and gadolinium isotopes by the PNC-CSM method. An gradual rise trend from $N=96$ to $102$ can be seen for all the states. The fermion dynamic symmetry model (FDSM) predicted that $B(E2)$ values will become saturated and start to bend down at $N>100$ due to the dynamical Pauli effect~\cite{WuC1994_ANP21}. However, according to the present PNC-CSM calculations, except for the $5^- \{\pi\frac{3}{2}[411]\otimes\pi\frac{7}{2}[523]\}$ state, no obvious bending down is exhibited at neutron number $N=100$. We note that the systematic calculations of the finite-range droplet model (FRDM) predicted a maximum quadruple deformation with $\varepsilon_{2}=0.275$ at neutron number $N=99\sim103$ for both of samarium and gadolinium neutron rich isotopes~\cite{MoellerP1995_ADaNDT59}. Accordingly, if the quadruple deformation parameters of M\"{o}ller \textit{et al.}~\cite{MoellerP1995_ADaNDT59} are used, it would lead to a clearly down-bending of $B(E2)$ at neutron number $N=100$ for all the states of samarium and gadolinium isotopes.  

In Fig.~\ref{fig:BE2_isotone}, it shows the $B(E2,\omega=0)$ value (in $e^2b^2$) systematics of the GSB along the $N=96,98,100,102$ isotone chains for neodymium, samarium, gadolinium and dysprosium by the PNC-CSM method. All the four isotone chains display the similar trend as proton number vary from $Z=60$ to $66$. The $B(E2)$ values increase sharply from neodymium to samarium since more valence nucleon participate in the collective behavior. It starts to saturate at $Z\geq62$ and a much gentle rise of $B(E2)$ is displayed from samarium to dysprosium. As the experimental moments of inertia for neodymium (not shown in the present paper) are higher than for samarium, the neodymium isotopes may have reduced pairing due to the energy gap at $Z=60$. Pairing of the ground state and low-lying multi-particle state bands in this mass region would be interesting for future study.  

The $B(E2)$ values as the function of rotation frequency $\hbar\omega$ of the GSB of neodymium, samarium, gadolinium and dysprosium isotopes are displayed in Fig.~\ref{fig:BE2}. A common feature can be seen that the $\omega$-dependent $B(E2)$ keeps almost constant at the low frequency, which reflects that the studied nuclei have the stable rotor character with large collectivity. As the frequency increasing, the $B(E2)$ values start bend down around $\hbar\omega>0.20$ MeV because of the anti-pairing Coriolis effect and the gradually increased alignment of the paired particles. From the left panel of Fig.~\ref{fig:BE2}, a slight bend down of $B(E2)$ at $N=100$ can be detected for neodymium, gadolinium and dysprosium isotopes, and it becomes comparatively clear as frequency increasing, especially for the neodymium isotopes.  

\section{Summary}{\label{Sec:summary}}
The high-$K$ isomeric states in neutron-rich even-even nuclei $^{158-164}$Sm and $^{160-166}$Gd have been studied by using the cranked shell model with the pairing treated by the particle-number conserving method. The experimental data including band head energies and moments of inertia are reproduced quite well by theoretical calculations. In most cases, the PNC-CSM calculations confirm the configuration assignments in the earlier works except for the $1.279$ MeV $5^-$ isomeric state in $^{158}$Sm, to which, both the state energy and the moment of inertia prefer the assignment of the two-proton $\pi\frac{5}{2}^{+}[413]\otimes\pi\frac{5}{2}^{-}[532]$ configuration by the PNC-CSM calculations. By analysis the moment of inertia, the 146 keV $\gamma$ ray in the spectrum of $^{162}$Sm is more likely to be the decay from a $6^{-}$ state to the $1.7\mu$s $4^-$ isomeric state. 

The high-order deformation $\varepsilon_{6}$ effect is nontrivial. It leads to the energy gaps at proton $Z=62,68$ and neutron $N=102$, and makes ones at proton $Z=60$ and neutron $N=98$ less pronouced. Accordingly, $20-450$ keV and $80-350$ keV variations in the multi-particle state energies are obtained compared to the $\varepsilon_{6}=0$ calculations for neutron and proton, respectively. In general, calculations with non-zero $\varepsilon_{6}$ result in a better reproduction of the experimental data.   

Possible low-lying two-particle states in samarium and gadolinium isotopes are predicted, especially for some systematic occurring states. These are the two-proton $5^-$ $\{\pi\frac{5}{2}^{+}[413]\otimes\pi\frac{5}{2}^{-}[532]\}$ and $4^-$ $\{\pi\frac{3}{2}^{+}[411]\otimes\pi\frac{5}{2}^{-}[532]\}$ states in samarium isotopes, and $4^+$ $\{\pi\frac{5}{2}^{+}[413]\otimes\pi\frac{3}{2}^{+}[411]\}$ state in gadolinium isotopes.  

The systematics of the electronic quadrupole transition probabilities $B(E2)$ values along the neodymium, samarium, gadolinium and dysprosium isotopes and $N=96,98,100,102$ isotones chains is investigated by the semiclassical approximation with the microscopic wave function being obtained by the PNC-CSM method. A gradual increase of the $B(E2)$ values from $N=96$ to $102$ can be seen for all the states in samarium and gadolinium isotopes. The predicted saturation of the $B(E2)$ values at $N=100$ is not clearly displayed in the present PNC-CSM calculations. The systematic behavior of the $B(E2)$ values from neodymium to dysprosium $(Z=96-102)$ shows that it saturates at $Z=62$ and a bend down appears at samarium isotopes. 

%\clearpage
\begin{acknowledgements}

One of the authors, Xiao-Tao He is grateful to Prof. P. Walker for his useful comments and carefully reading the manuscript. This work is supported by the National Natural Science Foundation of China (Grant Nos. 11775112 and 11275098).
\end{acknowledgements}

%% Text of bibliographic itemh
\bibliographystyle{apsrev4-1}
\bibliography{../../References/ReferencesXT}
%\bibliography{NeutronRichIsomer}

\end{document}